\documentclass[structabstract]{aa}  
\usepackage{natbib}
\bibpunct{(}{)}{;}{a}{}{,} 
\usepackage{graphicx}
\usepackage{txfonts}
\newcommand{\cD}{{\cal D}}
\newcommand{\Naverage}[1]{\left\langle #1\right\rangle_N}
\newcommand{\Saverage}[1]{\left\langle #1\right\rangle_S}
\newcommand{\average}[1]{\left\langle #1\right\rangle_\cD}
\newcommand{\BR}{\mathbb{R}}
\newcommand{\rmd}{\mathrm{d}}
\authorrunning{B. Kalus, D.J. Schwarz, M. Seikel \& A. Wiegand}

\begin{document}
   \title{Constraints on anisotropic cosmic expansion from supernovae}

   \subtitle{}

   \author{Benedict Kalus\inst{1}\thanks{\tt kalus at physik dot uni-bielefeld dot de},
   		Dominik J. Schwarz\inst{1}\thanks{\tt dschwarz at physik dot uni-bielefeld dot de},
   		Marina Seikel\inst{2,3}\thanks{\tt marina dot seikel at uct dot ac dot za}
		and 
		Alexander Wiegand\inst{4}\thanks{\tt alexander dot wiegand at aei dot mpg dot de}}

   \institute{Faculty of Physics, Bielefeld University, Postfach 100131, 33501 Bielefeld, Germany
   		\and 
   		Department of Physics, University of the Western Cape, Cape Town 7535, South 
		  Africa
		\and
		Astrophysics, Cosmology and Gravity Centre (ACGC), and Department of Mathematics 
		  and Applied Mathematics, University of Cape Town, Rondebosch, 7701, South Africa
		\and
		Max-Planck-Institut f\"ur Gravitationsphysik, Albert-Einstein-Institut, Am M\"uhlenberg 1, 14476 Potsdam, Germany}

   \date{Received ; accepted }

 
  \abstract
   {}
   {We test the isotropy of the expansion of the Universe by estimating the hemispherical 
   anisotropy of supernova type Ia (SN Ia) Hubble diagrams at low redshifts ($z<0.2$).}
   {We compare the best fit Hubble diagrams in pairs of hemispheres and search for the maximal asymmetric 
   orientation. For an isotropic Universe, we expect only a small asymmetry due to noise and the presence of
   nearby structures. This test does not depend on the assumed content of the Universe, the assumed model 
   of gravity, or the spatial curvature of the Universe. 
   The expectation for possible fluctuations due to large scale structure is evaluated for the $\Lambda$ cold dark matter ($\Lambda$CDM)
   model and is compared to the 
   supernova data from the Constitution set for four different light curve fitters, thus allowing 
   a study of the systematic effects.}
   {The expected order of magnitude of the hemispherical asymmetry of the Hubble expansion 
   agrees with the observed one. The direction of the Hubble asymmetry is established 
   at 95\% confidence level (C.L.) using both, the MLCS2k2 and the SALT II light curve fitter. The highest expansion rate is found towards 
   $(\ell, b)\approx (-35\degr,-19\degr)$, which agrees with directions reported 
   by other studies. Its amplitude is not in contradiction to expectations from the $\Lambda$CDM model. The 
   measured Hubble anisotropy is $\Delta H/H \sim 0.026$. With 95\% C.L. the expansion asymmetry is $\Delta H/H<0.038$.}
   {}


   \maketitle
%

\section{Introduction}
\label{sec:intro}

Isotropy of the distributions of matter and light in the Universe is 
one of the basic assumptions of modern cosmology. The smallness of the temperature anisotropies 
in the cosmic microwave sky shows that this assumption is an excellent approximation. However, the 
cosmic microwave sky probes the symmetry of the Universe at the time of photon decoupling and it 
is interesting to test the assumed isotropy at smaller redshifts. A recent study of various 
probes at much smaller redshift reveals fluctuations at the per cent level and larger \citep{Gibelyou}.
These can partly be explained by local large scale structure, partly by systematics of the observations,
but some of them might be at odds with our understanding of cosmology and call for further investigation.

Complementing the assumption of isotropy with the additional assumption of homogeneity 
predicts the space-time metric to become of the Robertson-Walker type, predicts the redshift of light 
$z$, and predicts the Hubble expansion of the Universe. Then the cosmic luminosity distance-redshift 
relation for comoving observers and sources becomes  
\begin{equation}
	d_L(z) = \frac{cz}{H_0}\left[1+\left(1-q_0\right)\frac{z}{2}\right]+\mathcal{O}\left(z^3\right)\, ,
	\label{eq:Hubble_law}
\end{equation}
with $H_0$ and $q_0$ denoting the Hubble and deceleration parameters, respectively. Note that this 
prediction holds for arbitrary spatial curvature, any theory of gravity (as long as space-time is described 
by a single metric) and arbitrary matter content of the Universe.

Assuming the cosmological principle, \cite{Seikel} showed that the acceleration of the cosmic expansion can be established in a 
model independent test (any curvature, any matter content, any dynamics of gravity) at 4.2 sigma from the 
Union data set of supernovae. One of the assumptions of this test is that supernovae are a fair sample of
isotropically distributed standard candles. In \cite{Seikel} it was demonstrated that local SNe ($z < 0.2$) are  
crucial to this conclusion. These local SNe come from a tiny fraction of the Hubble volume only, thus it 
is not obvious, if they form a fair sample and if they are sampled from an isotropic and homogeneous 
distribution. Testing this assumption is the purpose of this work. 

A simple test for isotropy relies on comparing Hubble diagrams for pairs of hemispheres. Statistically 
significant deviations from an isotropic Hubble diagram have been discovered by \citet{Schwarz} 
at redshifts $z<0.2$ in three different data sets. However, the direction of maximal hemispheric asymmetry is close to the equatorial poles, and thus a
systematic error in the SN search, observation, analysis or data reduction seems to be plausible.

A more natural direction of anisotropy was identified by \citet{Colin}. They found out that the low redshift regime ($z<0.06$) 
of the data in the Union 2 catalogue \citep{Amanullah} is barely consistent with $\Lambda$CDM at $2-3\;\sigma$ 
because of a bulk flow towards the Shapley supercluster.

\citet{Antoniou} also applied the hemisphere comparison method to the SN Ia data of Union 2 and found a maximum anisotropy, which itself 
is statistically not significant, but coincides with the bulk velocity flow axis \citep{Watkins, Kashlinsky, Lavaux}, cosmic microwave background (CMB) low 
multipole moments \citep{Bennett, Copi} and the quasar optical 
polarisation alignment axis \citep{Hutsemekers}. Similar directions can be obtained by considering a dynamical dark energy fluid with
an anisotropic equation of state in the wCDM model or in Chevallier-Polarski-Linder parametrisation \citep{Cai}. {SNe Ia have also been used by 
\citet{Koivisto07, Koivisto08} to test other models of anisotropic dark energy and by \citet{Koivisto10} to constrain anisotropic curvature.} 
Recently, \citet{Mariano} detected an alignment of supernova dipole asymmetries and a dipole of a locally varying fine structure constant, 
which can be observed in the Keck+VLT quasar absorber sample. {\citet{Jackson} found a hemispherical anisotropy in the matter density $\Omega_m$ at higher redshifts 
($0.5<z\leq 3.787$), where the smallest value of  $\Omega_m$ points close towards the CMB dipole. An additional tool to test the assumption of isotropy consists of
measuring polarisations of different astrophysical sources, as anisotropic expansion has effects in electrodynamics leading to observational consequences
\citep{Ciarcelluti}.}

In the present paper, we analyse directions of maximally asymmetric Hubble expansion constraining its maximum value. Besides, we infer from SN data typical 
fluctuations of the expansion rate and compare them with expectations for the $\Lambda$CDM model. The use of SN data fitted with 4 different light-curve fitters
by \citet{Hicken} allows us to study systematic effects.

The paper is structured as follows: We explain the method of our test and state the corresponding theoretical expectation on the fluctuations
in section \ref{sec:hemasym}. We continue with describing the data set, which we will analyse in section \ref{sec:data}.
Statistical analyses can be found in section \ref{sec:statistics} and possible systematics will be discussed in section \ref{sec:systematics}. 
Upper and lower limits on the expansion asymmetry are given in section \ref{sec:results}. We conclude in section \ref{sec:concl}.

\section{Hemispherical asymmetries} 
\label{sec:hemasym}

Our aim is to test whether the Hubble diagram is isotropic. In order to do so, we apply a simple test, that first 
has been used by \citet{Eriksen} to study asymmetries in the CMB anisotropy field, and that \citet{Schwarz} 
adopted to find anisotropies in SN data. We test the anisotropic cosmic expansion by splitting the sky into 
hemispheres. Having calculated the Hubble diagrams on both hemispheres separately, one can change the 
directions of the poles and recalculate the Hubble diagrams on the new hemispheres.

We fit the Hubble diagrams by applying the method of least squares to the distribution of distance moduli of the supernovae, 
using as a model the relation of the distance modulus to the redshift given by (\ref{eq:Hubble_law}) and the definition
\begin{equation}
	\mu=5\log_{10}\left(d_L/\mathrm{Mpc}\right)+25\, .
	\label{eq:dist_mod}
\end{equation}
\citet{Schwarz} estimated, that the error from truncating Hubble's law as in (\ref{eq:Hubble_law}) is
below 10\% if one fits only at redshifts $z<0.2$. Despite this restriction, this test has the major advantage, 
that it is model-independent, because (\ref{eq:Hubble_law}) follows directly from the cosmological 
principle and does not contain any terms depending on the curvature, theory of gravity or content of the Universe. 

{Admittedly, as to go from SN magnitudes $m=\mu+{\cal M}$ to the luminosity distance $d_L$, one has to know the nuisance parameter 
${\cal M}\equiv M-5\log_{10}\left(H_0\right)+25$, which also depends on the absolute magnitude $M$ and hence on the physics of SNe
\citep{Perlmutter}. We assume that Type Ia SNe are standardisable, thus we treat $\cal M$ as a universal number and use the distance moduli which \citet{Hicken} 
calculated by
using four different light-curve fitters with calibrations $H_0=65$ km/s/Mpc and $M_B=-19.46$ for SALT, $M_B=-19.44$ for SALT II and $M_V=-19.504$ for MLCS2k2.}

{When fitting the Hubble diagrams, we allow $H_0$ to vary from hemisphere to 
hemisphere, but keep the deceleration parameter at its best-fit $\Lambda$CDM model value. 
The SN Hubble diagrams considered here do not constrain $q_0$ well enough, to justify the 
inclusion of this parameter in the fit.}

In a perfectly homogeneous and isotropic Universe we would expect any nontrivial result in the 
hemispherical asymmetry test to be due to noise and systematics. However, the large scale structure of the 
Universe is expected to give rise to a small deviation from isotropy. We can estimate the expected 
amount of fluctuations of the Hubble rate between the hemispheres by estimating their cosmic variance
\citep{Turner, Shi, Wang, Li, Wiegand}. To measure the difference between the two hemispheres we use the quantity
$\frac{H_{N}-H_{S}}{H_{N}+H_{S}}$. In the formalism of \cite{Wiegand} the Hubble
rate on each of the hemispheres is given by the volume average over
the local expansion rates. In the linear regime, these are related
to the density contrast and therefore the average Hubble rate on a
hemisphere is 
\begin{equation}
H_{S/N}=\overline{H_{S/N}}\left(1-\frac{1}{3}f_{\cD_{0}}\left\langle \delta_{0} \right\rangle_{S/N}\right)\;.
\end{equation}
where the growth rate factor today $f_{\cD_{0}}\approx0.5$ in the
standard $\Lambda$CDM model. To first order the anisotropy therefore
is 
\begin{equation}
\frac{H_{N}-H_{S}}{H_{N}+H_{S}}=\frac{1}{6}f_{\cD_{0}}\left(\left\langle \delta_{0} \right\rangle_{S} - \left\langle \delta_{0} \right\rangle_{N} \right)\label{eq:anisofo}
\end{equation}
So the ensemble expectation value of the anisotropy is zero at leading
order. However, we are interested in the typical fluctuations between
different locations in the Universe. Using (\ref{eq:anisofo}) we
find 
\begin{equation}
\sigma_{A}=\sigma\left(\frac{H_{N}-H_{S}}{H_{N}+H_{S}}\right)=\mathbf{\frac{1}{3}f_{\cD_{0}}\sqrt{\sigma_{HS}^{2}-\sigma_{FS}^{2}}}\label{eq:flucaniso}
\end{equation}

{where}
\begin{equation}
\sigma_{HS}^{2}:=\sigma^{2}\left(\Naverage{\delta_{0}}\right)=\sigma^{2}\left(\Saverage{\delta_{0}}\right)\label{eq:defsighs}
\end{equation}
{is the matter variance on a hemisphere and}
\begin{equation}
\sigma_{FS}^{2}:=\sigma^{2}\left(\average{\delta_{0}}\right)=\int_{\BR^{3}}P_{0}\left(k\right)\widetilde{W}_{\cD}^{2}\left(k\right)\rmd^{3}k\;.\label{eq:defsigfull}
\end{equation}
{is the matter variance on the full sphere. $P_{0}\left(k\right)$
is the linear matter power spectrum and $\widetilde{W}_{\cD}\left(k\right)$
is the Fourier transform of the radial window function $W_{\cD}\left(r\right)$
of the domain $\cD$. This window function is often taken to be of
top-hat-- or Gaussian--form. To arrive at more precise predictions,
however, we shall use in Sec. \ref{sec:data} a shape that describes the actual
SN distribution more accurately.}

{The explicit form of $\sigma_{HS}^{2}$ is more complicated than the
one for the full sphere, because a window function with the geometry
of a hemisphere is no longer spherically symmetric. Therefore, the
window function has also an angular part in addition to the radial
part $W_{\cD}\left(r\right)$. We calculate this angular part using
a decomposition into spherical harmonics (see appendix A of \cite{wiegand2012}
for
the explicit form). Apart from this angular window function, which
is fixed by the requirement that we consider a hemisphere, $\sigma_{HS}^{2}$
also uniquely depends on $P_{0}\left(k\right)$ and $W_{\cD}\left(r\right)$.
Thus, using the fitting formula of \cite{eisensteinhupower} for the matter power
spectrum
allows us to estimate the possible anisotropies in a $\Lambda$CDM
universe in Sec. \ref{sec:data}.}

\section{Supernova data and results}
\label{sec:data}

For our analysis, we use the Constitution set \citep{Hicken} supplemented 
by the positions of the SNe procured from the list of SNe 
\footnote{\texttt{http://www.cbat.eps.harvard.edu/lists/Supernovae.html}}
provided by the 
IAU Central Bureau for Astronomical Telegrams (CBAT). The Constitution
set contains a large number of nearby SNe from all the sky (except for the zone of avoidance), whose distance moduli
have been obtained by four different light-curve fitters: Multicolor Light Curve Shapes 2k2 (MLCS2k2) (1.7),
MLCS2k2(3.1), Spectral Adaptive Light curve Template (SALT) and SALT II. Thus, the Constitution set consists of
four samples. 

The SALT \citep{Guy05} fitter describes light-curves by the peak
apparent $B$-band magnitude $m_B^{max}$, the stretch factor $s$
and the color parameter $c$. SALT does not distinguish between the
intrinsic color of a SN and the reddening which is caused by dust
extinction. Both effects are summarised in $c$, thus assuming the same
color-magnitude relation for the combination of color variation and
host reddening at low and at high redshift.
SALT II \citep{Guy07} is similar to SALT, but uses a different
spectral template. $s$ has been replaced by a new stretch parameter
$x_1$.

In contrast to the SALT fitters, MLCS2k2 \citep{Jha} includes the
intrinsic color in the shape parameter $\Delta$ assuming a
broader-bluer and narrower-redder relation. The host-galaxy extinction
parameter $A_v$ is then determined separately. This requires
assumptions about the dust properties specified by the reddening
parameter $R_V$. In the Milky Way, this parameter has been measured to
be $R_V=3.1$. We will refer to the MLCS2k2 fitter using this value as
MLCS2k2(3.1). \cite{Hicken} have determined the reddening parameter,
which minimises the scatter in the Hubble diagram, to be
$R_V=1.7$. In the following, the fitter with this value will be
denoted as MLCS2k2(1.7).
 
\begin{figure}
	\includegraphics[width=\linewidth]{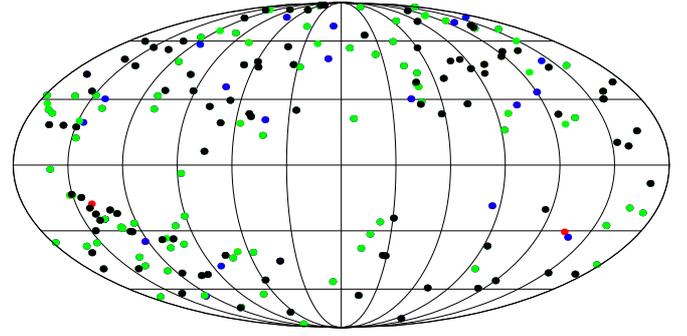}
	\caption{Celestial distribution of SNe from \citet{Hicken} having redshifts $z<0.2$ in galactic coordinates. Black dots represent SNe, that are in all four 
	  samples, while green ones 
	  cannot be found in the SALT sample, blue ones are only in the MLCS2k2 samples. The MLCS2k2 ($R_V=3.1$) sample is the only subset, which contains red SNe. We 
	  plot in Mollweide projection.}
	\label{fig:dist_SN_on_Sky_4_all_samples}
\end{figure}

While there is a significant overlap of the SNe
contained in the samples, not all observed SNe can be found in a
specific sample. SNe are rejected due to certain criteria, e.g. if the
light-curve fit is not satisfactory or if the light extinction is too
high. The different parameters and rejection criteria of the fitters
thus lead to a differing number of SNe in the samples.
Fig.~\ref{fig:dist_SN_on_Sky_4_all_samples} shows the distribution of
SNe on the sky for all four samples. In Table \ref{tab:means} the
number of SNe with $z<0.2$, their mean and median redshifts and their
mean $\sigma_\mu$ are listed for all four samples. The SALT sample
contains a significantly smaller number of SNe than the other three
samples as the stricter rejection criteria from \cite{Kowalski} have
been adopted.

\begin{table}
	\[
		\begin{array}{p{0.25\linewidth}|*{4}{p{0.16\linewidth}|}}
			\hline
			\noalign{\smallskip}
			fitter & \# SN with $z<0.2$ & mean \mbox{redshift} & median redshift & mean $\sigma_\mu$ \mbox{in mag}\\
            \noalign{\smallskip}
            \hline
            \noalign{\smallskip}
            MLCS2k2 (1.7) & 199 & 0.030 & 0.025 & 0.191\\ 
            MLCS2k2 (3.1) & 203 & 0.030 & 0.025 & 0.197\\
            SALT & 115 & 0.037 & 0.028 & 0.187\\
            SALT II & 183 & 0.031 & 0.026 & 0.218\\
            \noalign{\smallskip}
            \hline
         \end{array}
    \]
	\caption{Compilation of some characteristics of the samples used throughout this article.}
	\label{tab:means}
\end{table}

\begin{figure}
	\centering
	\includegraphics[width=\linewidth]{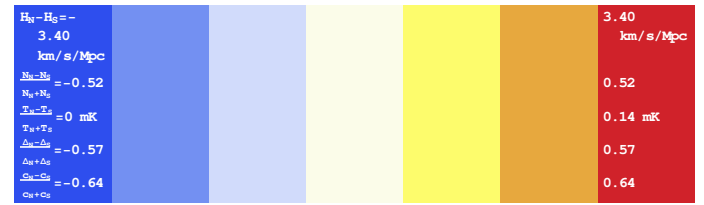}
	\caption{Colour scaling of all maps drawn for this article. $H_N-H_S$ denotes the difference in the expansion rates for a pair of hemispheres, 
	  $\frac{N_N-N_S}{N_N+N_S}$ is the corresponding number asymmetry of SNe, $\frac{T_N-T_S}{T_N+T_S}$ is the asymmetry in dust antenna temperature,
	  $\frac{\Delta_N-\Delta_S}{\Delta_N+\Delta_S}$ stands for the asymmetry in the shape parameter of the MLCS2k2 fitter,
	  and $\frac{c_N-c_S}{c_N+c_S}$ is the colour parameter asymmetry (SALT/SALT II).}
	\label{fig:Scale}
\end{figure}

\begin{figure*}
	\includegraphics[width=0.5\textwidth]{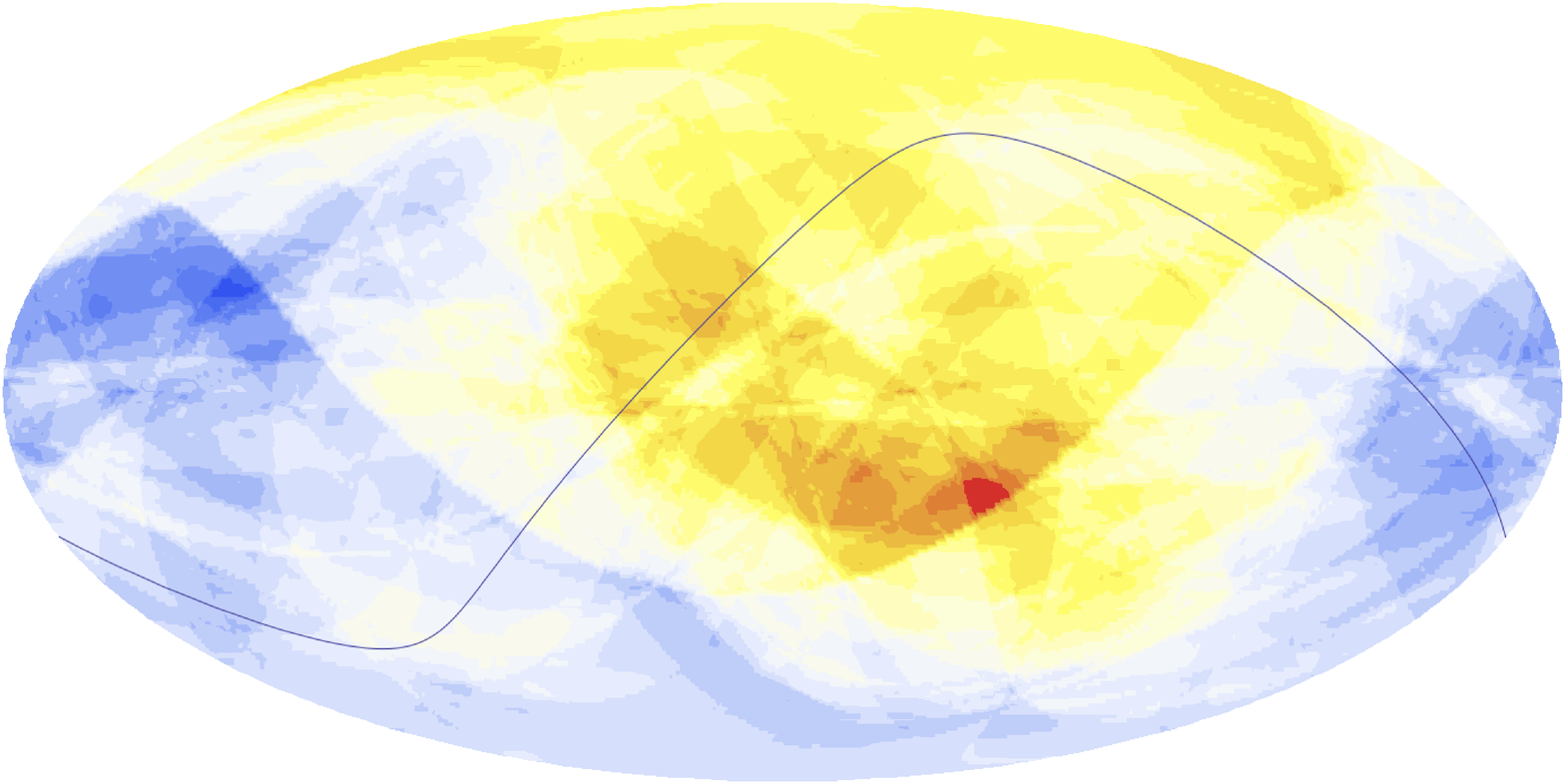}
	\includegraphics[width=0.5\textwidth]{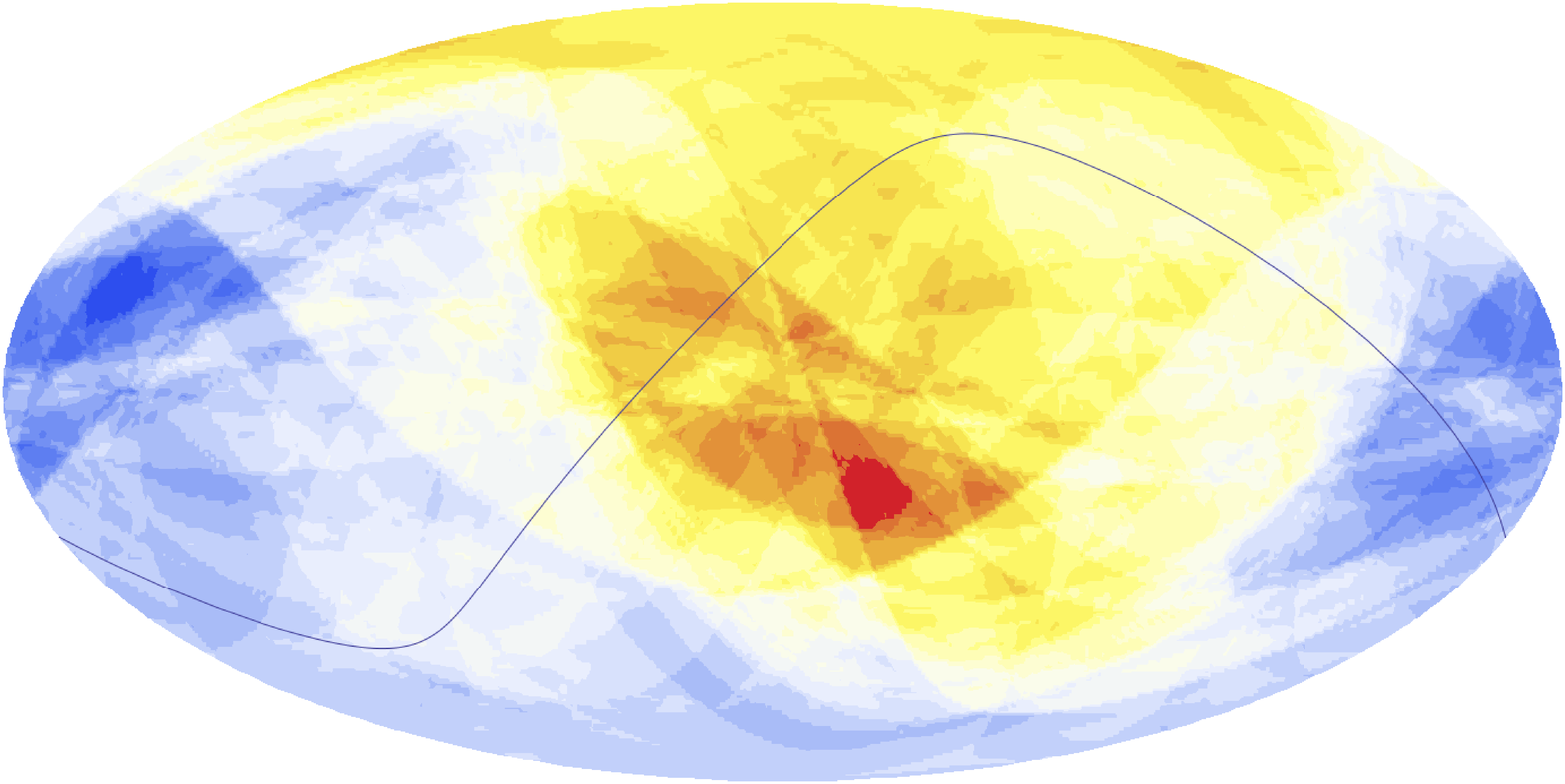}
	\includegraphics[width=0.5\textwidth]{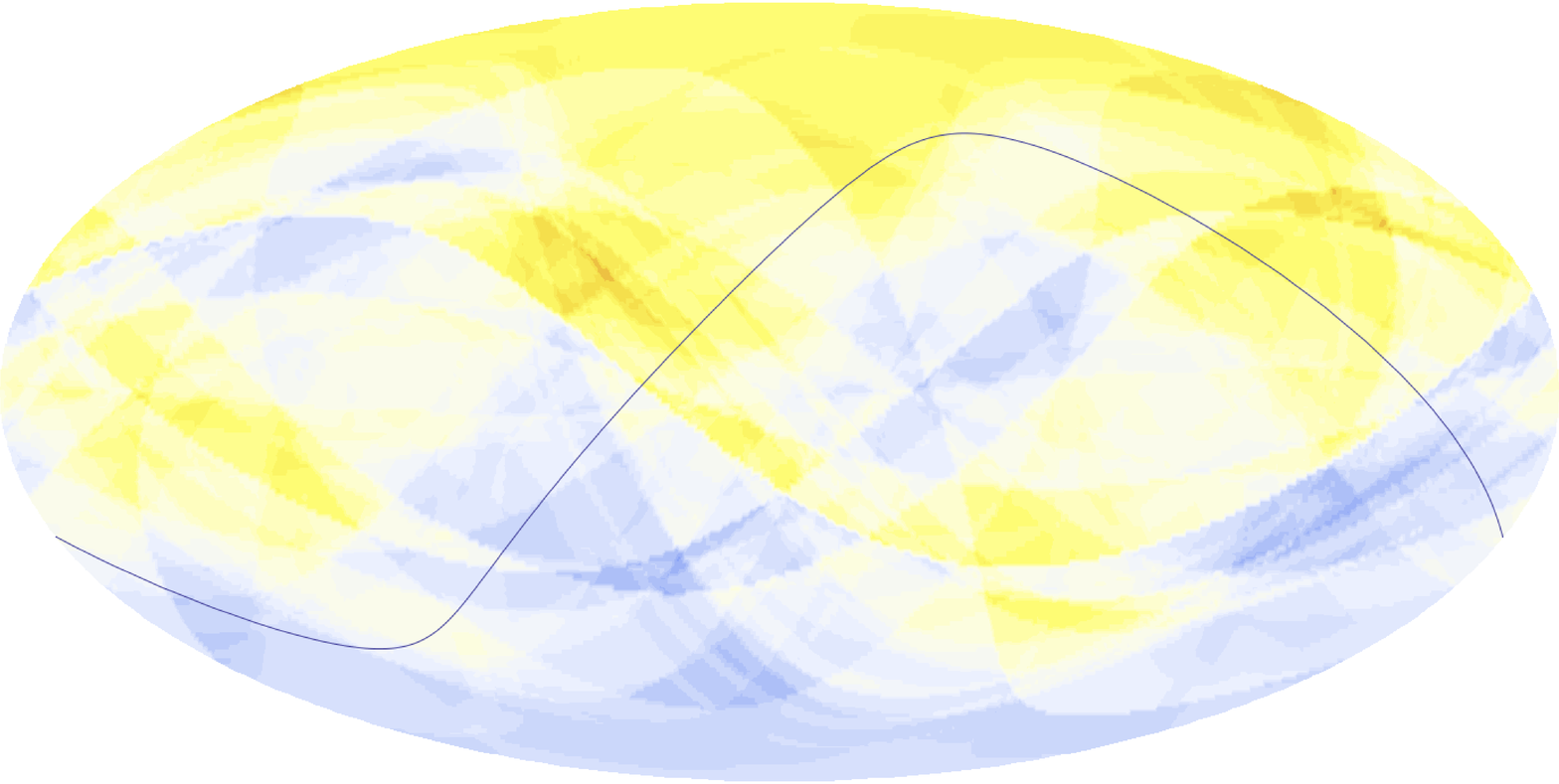}
	\includegraphics[width=0.5\textwidth]{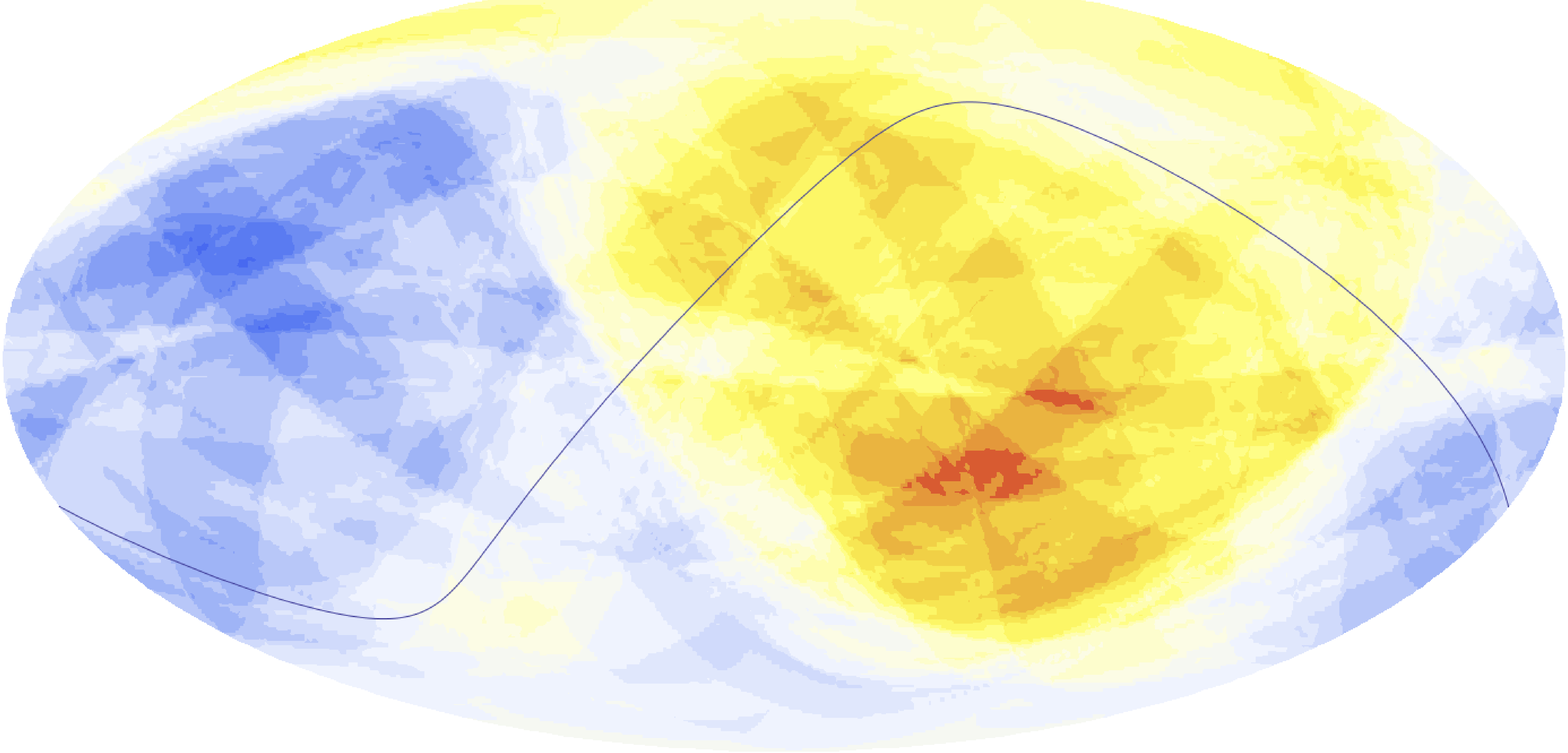}
	\caption{Hemispherical asymmetry $H_N-H_S$ in the Hubble rate $H_0$ for each SN Ia fitter. The 
	   upper left plot is for MLCS2k2 (1.7), the upper right one is for MLCS2k2 (3.1), below are SALT (left) and SALT II (right). 
	   The deceleration parameter is fixed at the value $q_0=-0.601$, which one obtains from WMAP \citep{Larson}. When 
	   the north pole lies in a region with a bluish colour, $H_N-H_S$ is negative, the red spots denote the directions 
	   of positive large asymmetry $H_N-H_S$. The Hubble parameter has no anisotropy pointing towards white areas. All 
	   plots showing the asymmetry in the Hubble rate have the same scaling. The scale is given in Fig. \ref{fig:Scale}. 
	   At the reddest and bluest spots, $\left\vert H_N-H_S\right\vert=3.4$. We plot directions in galactic coordinates 
	   and Mollweide projection. The black line is the equator of the equatorial coordinate system.}
	\label{fig:HN-HS_while_q0_fixed}
\end{figure*}

In our data analysis, we set the deceleration parameter $q_0=-\Omega_\Lambda+\frac{1}{2}\Omega_M=-0.601$ according to the 
seven-year Wilkinson Microwave Anisotropy Probe (WMAP) observations fitted to $\Lambda$CDM \citep{Larson}.

{The effect of our choice of $q_0$ is not too large: Repeating our test described below for SALT II
 and fixing $q_0$ at different values between -1.5 and 0.4, yields values of 
$\left(H_N-H_S\right)_{\max}$ which vary almost linearly between 3.20 km/s/Mpc for $q_0=-1.5$
and $\left(H_N-H_S\right)_{\max}=2.55$ km/s/Mpc for $q_0=0.4$.}

{As said in section \ref{sec:intro}, we carry out a $\chi^2$-fit to adjust the Hubble parameters $H_{N}(\ell,b)$ on the hemisphere identified by its pole $(\ell, b)$ 
and $H_S(\ell, b)$ on the complementary hemisphere such that the distance moduli modelled by (\ref{eq:dist_mod}) and related to redshift data by (\ref{eq:Hubble_law})
fits the distance moduli given in \citet{Hicken} best.}
We divide the sky into a $1^\circ\times 1^\circ$ grid. Each grid point can be regarded as the pole $(\ell, b)$ of a hemisphere.
We determine the Hubble parameter $H_N(\ell,b)$ and subtract from it the Hubble parameter $H_S(\ell,b)$ of the opposite side. 
The resulting deviations in $H_0$ are plotted for each light-curve fitter in Fig. \ref{fig:HN-HS_while_q0_fixed}.

The asymmetry maps for the SALT fitter exhibits several small local extrema in $H_N-H_S$, which look like noise, while the maps obtained with the MLCS2k2 fitters and 
SALT II are similar to each other: The white areas, which denote small deviations in the Hubble parameter, have similar shapes, and the directions of maximum asymmetry
are also similar, as can be verified in Table \ref{tab:Results_q0_fixed}. Table \ref{tab:Results_q0_fixed} also shows, that the maxima of 
$\frac{H_N-H_S}{H_N+H_S}$ for the MLCS2k2 fitters and SALT II are well above 2\% and hence larger than the fluctuations $\sigma_A$ due to cosmic variance (cf.
 end of 
this section). MLCS2k2(1.7) and SALT II yield almost the same directions of maximum asymmetry. The maximally asymmetric direction in the MLCS2k2(3.1) is located in the
same quadrant as the previous ones, whilst the direction found in the SALT data is far from the others. One finds also a much smaller value of 
$\frac{H_N-H_S}{H_N+H_S}=1.54\%$ using the SALT data.

\begin{figure*}
\includegraphics[width=0.47\linewidth]{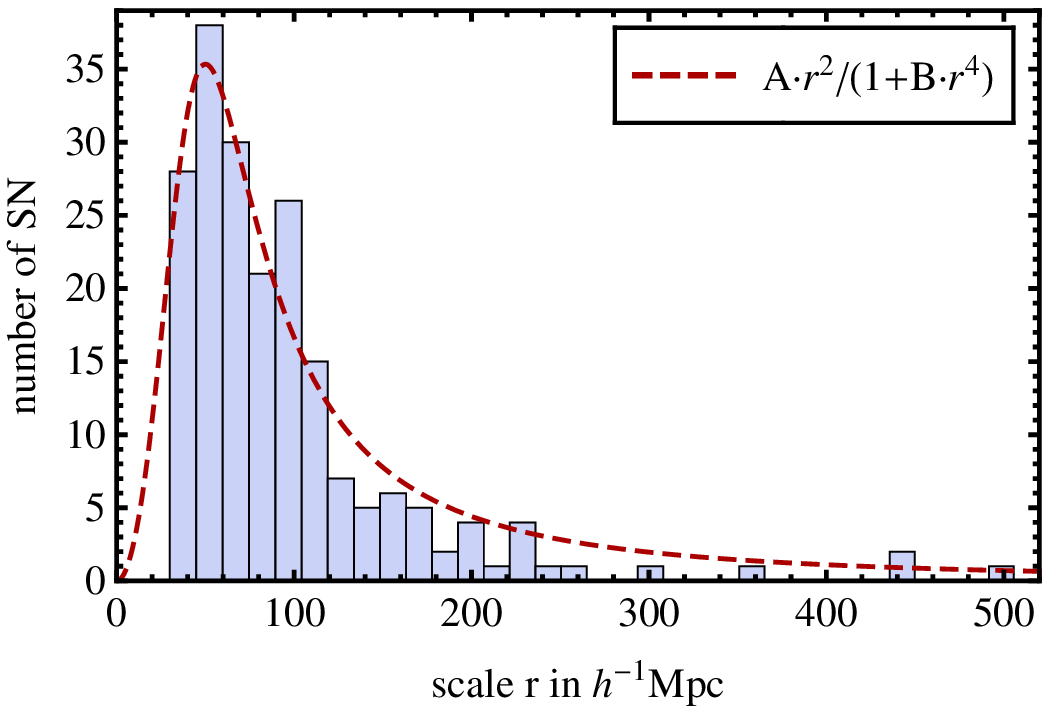}
\hspace{0.06\linewidth}
\includegraphics[width=0.47\linewidth]{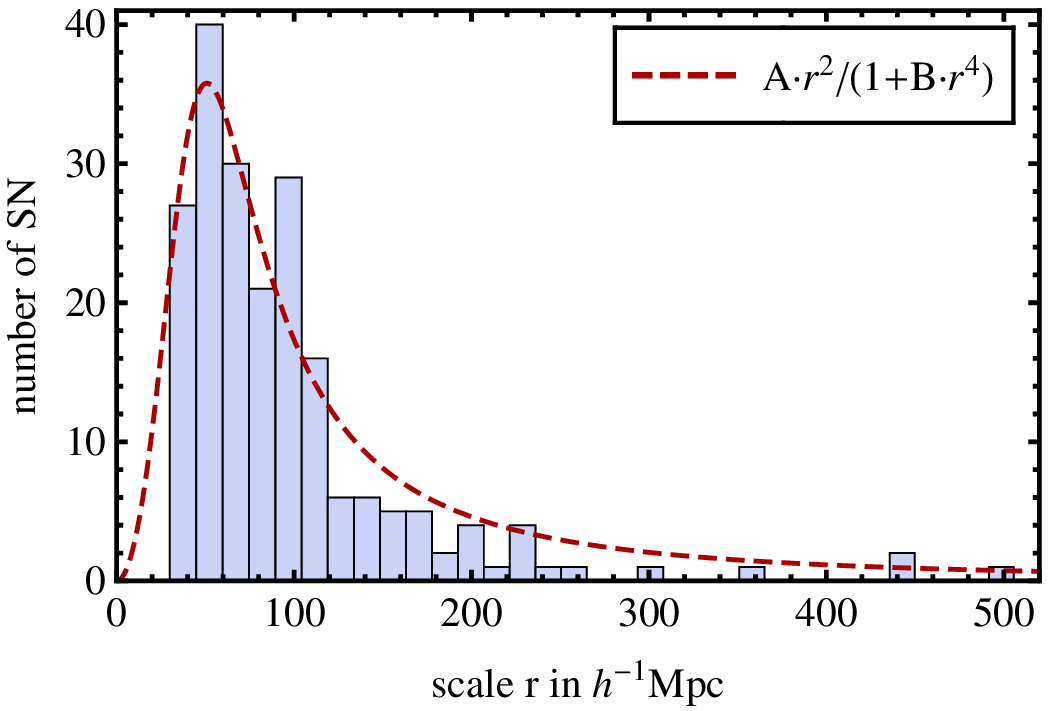}

\includegraphics[width=0.47\linewidth]{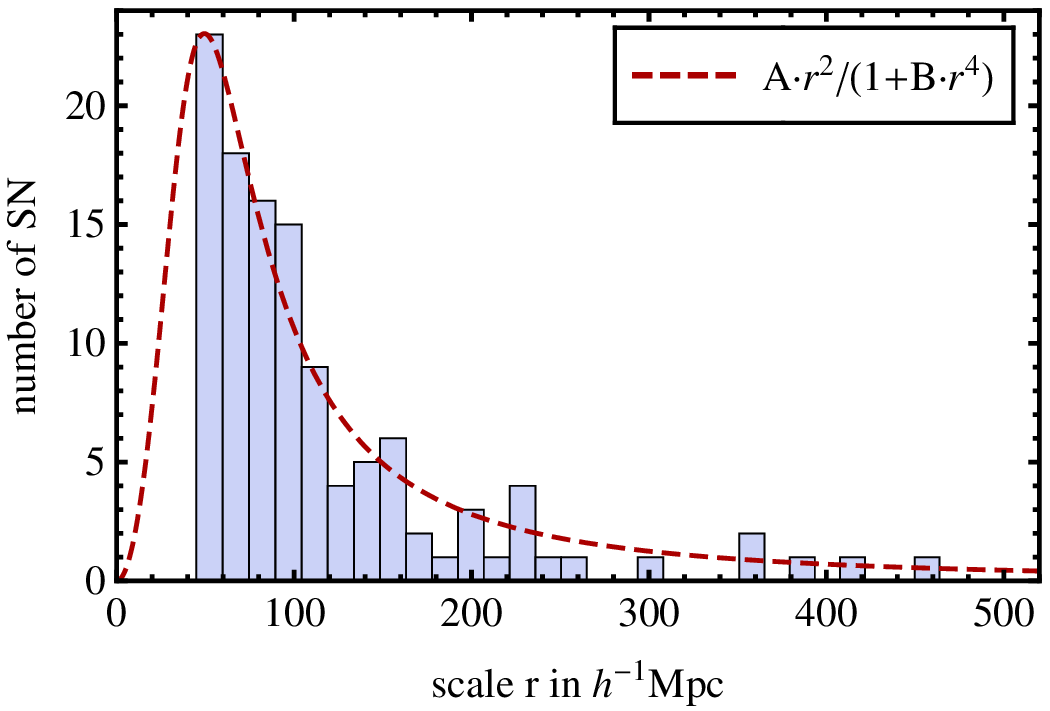}
\hspace{0.06\linewidth}
\includegraphics[width=0.47\linewidth]{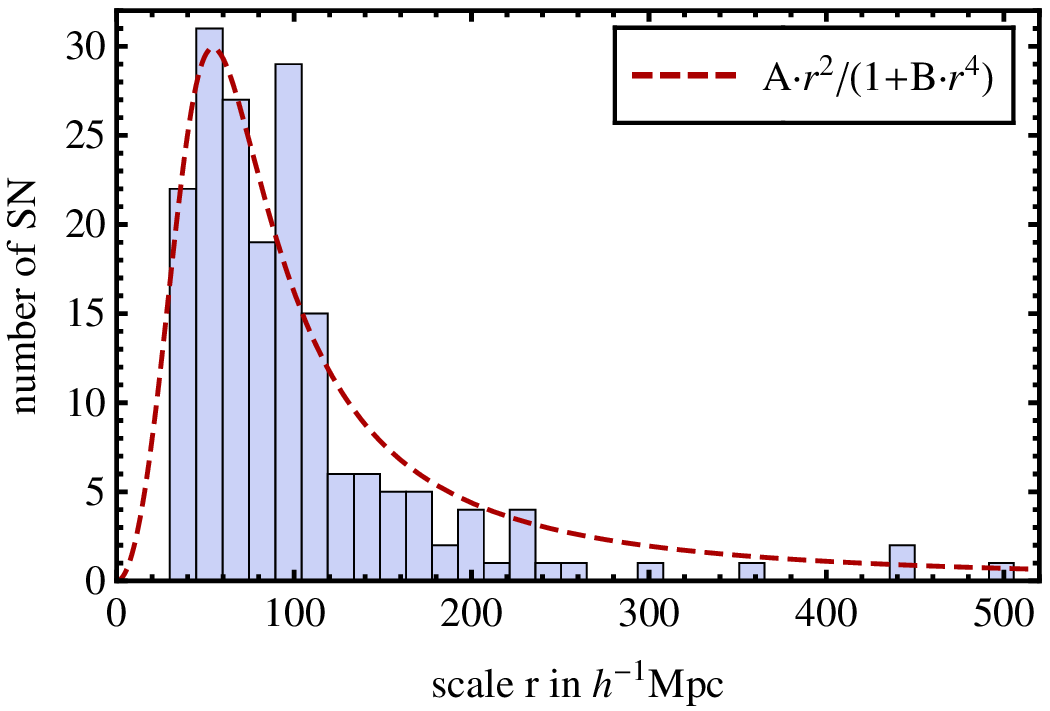}

\caption{The radial distribution of the supernovae in the MLCS2k2 (1.7) (top left), MLCS2k2 (3.1) (top right), SALT (bottom left) and SALT II (bottom right) data set. 
  The red dashed lines illustrate the radial window functions which we use in our calculations for the expected fluctuations.
\label{fig:RedshiftDistrib}}

\end{figure*}

{
For a correct interpretation of these anisotropies, we need to compare
them to what we expect by cosmic variance. In a statistically homogeneous
and isotropic $\Lambda$CDM model, the average hemispherical
asymmetry is zero. A particular realisation, however, will have an
asymmetry. The broadness of the distribution of these asymmetries
around zero, is the cosmic variance for the hemispherical asymmetry.
We use (\ref{eq:flucaniso}) to estimate it for the
four samples. To arrive at reliable predictions for the expected fluctuations
in the anisotropy, we have to account for the incompleteness of the
sampling of the hemispheres by the supernovae. To this end, we use
a radial window function that is adapted to the shape of the redshift
distribution of the SNe. We show this redshift distribution in Fig.
\ref{fig:RedshiftDistrib} together with the results
of a fit with the ansatz $Ar^{2}/\left(1+Br^{4}\right)$.
We use this shape as the radial window function
$W_{\cD}\left(r\right)\propto A/\left(1+Br^{4}\right)$
in the evaluation of (\ref{eq:defsighs}) and (\ref{eq:defsigfull}).}

{Plugging the resulting values for $\sigma_{HS}^{2}$ and $\sigma_{FS}^{2}$
into (\ref{eq:flucaniso}), we therefore arrive at a value of
about $0.7\%$ as shown in Table~\ref{tab:meandeltaH0}.
Note that this is the expectation value for a randomly fixed direction,
not for the direction of maximum asymmetry. We checked if the shape
of the radial window function is important by using different forms,
including a simple interpolation of the distribution of SN. The values
varied between $0.6\%$ and $1.0\%$
indicating that, in any case, the cosmic variance is well below $2\%$.
}

On the other hand, Table~\ref{tab:means} tells us, that the mean measurement uncertainty $\sigma_\mu$ of the distance modulus is approximately $0.2~\mathrm{mag}$ and 
that the number of SNe with $z<0.2$ is around 200 for the MLCS2k2 data. Thus, we calculate the Hubble parameters with roughly 100 objects per hemisphere, reducing the 
overall uncertainty in $\mu$ to an approximate value of $0.02$, which relates to an uncertainty in the Hubble parameter of the order of 0.01.

Hence, the sensitivity of this test applied to the Constitution set is at the same order as the expected fluctuations of $\Lambda$CDM, which we therefore cannot 
see. However, more extreme models would stick out.

\begin{table}
	\begin{tabular}{p{0.24\linewidth}p{0.09\linewidth}p{0.09\linewidth}p{0.21\linewidth}p{0.09\linewidth}p{0.09\linewidth}}
		\hline
		\noalign{\smallskip}
		fitter & $\ell$ & $b$ & $\left(H_N-H_S\right)_{max}$ & $\frac{H_N-H_S}{H_N+H_S}$\\
		\noalign{\smallskip}
		\hline
		\noalign{\smallskip}
		MLCS2k2 (1.7) & -52\degr & -19\degr & 3.3 km/s/Mpc & 2.49\%  \\
		MLCS2k2 (3.1) & -18\degr & -20\degr & 3.4 km/s/Mpc & 2.59\%  \\
		SALT & -154\degr & 32\degr & 2.0 km/s/Mpc & 1.54\%  \\
		SALT II & -52\degr & -18\degr & 3.0 km/s/Mpc & 2.28\%  \\
		\noalign{\smallskip}
		\hline
	\end{tabular}	
	\caption{Galactic longitudes $\ell$ and latitudes $b$ of the directions of maximum asymmetry $\left(H_N-H_S\right)_{max}$ in the Hubble rate $H_0$ if the 
	  deceleration parameter $q_0$ is fixed.}
	\label{tab:Results_q0_fixed}
\end{table}

\section{Statistical analysis}
\label{sec:statistics}

Having found asymmetries, we want to focus on the question of the likeliness of the maximal asymmetry observed on the sky. 
Therefore, we performed the following analyses:

\subsection{Scrambled data}

\begin{figure*}
	\includegraphics[width=0.47\linewidth]{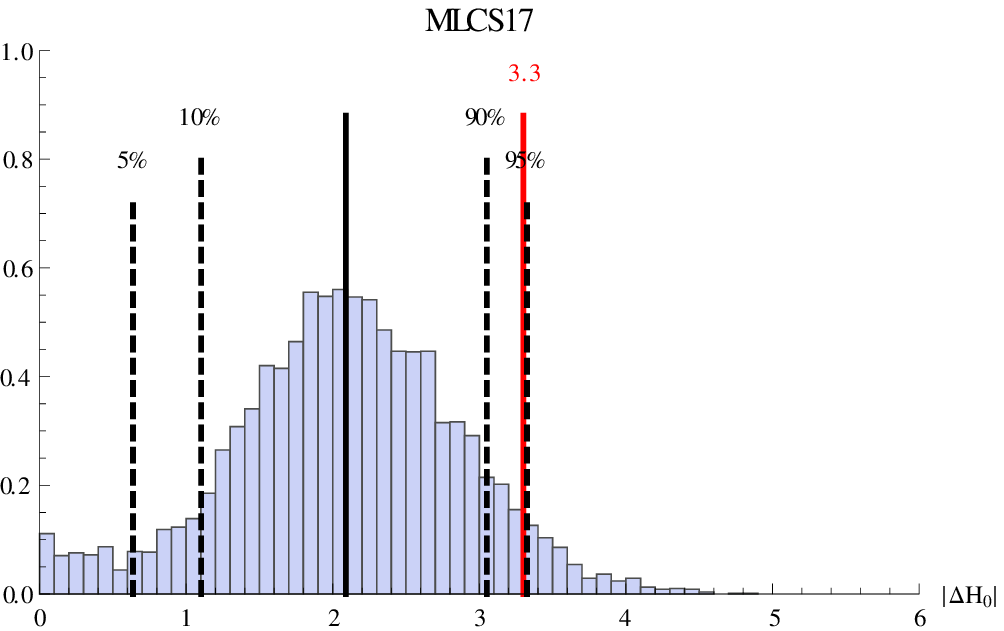}
	\hspace{0.06\linewidth}
	\includegraphics[width=0.47\linewidth]{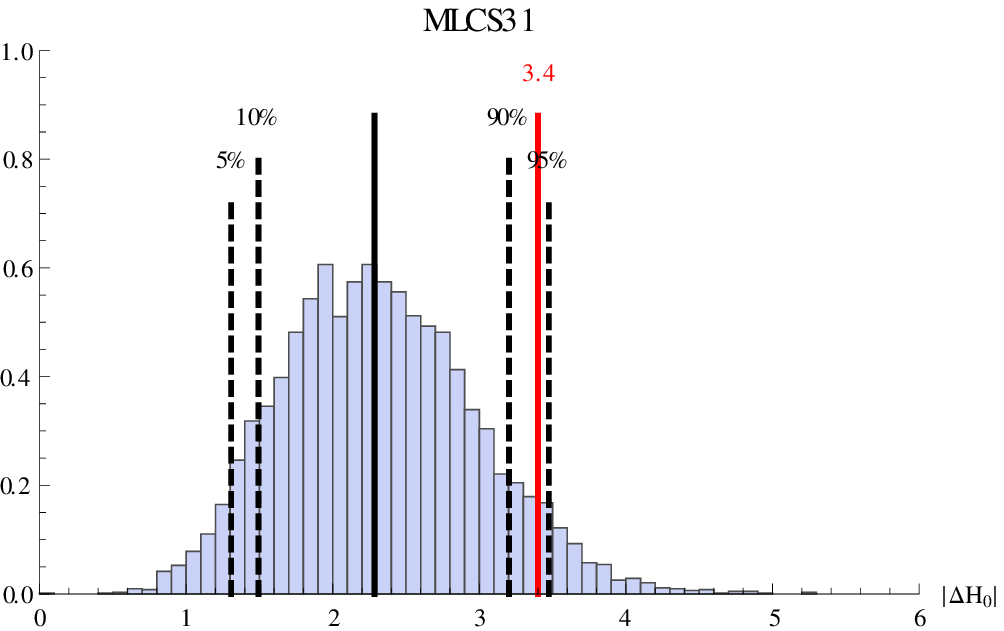}
	\includegraphics[width=0.47\linewidth]{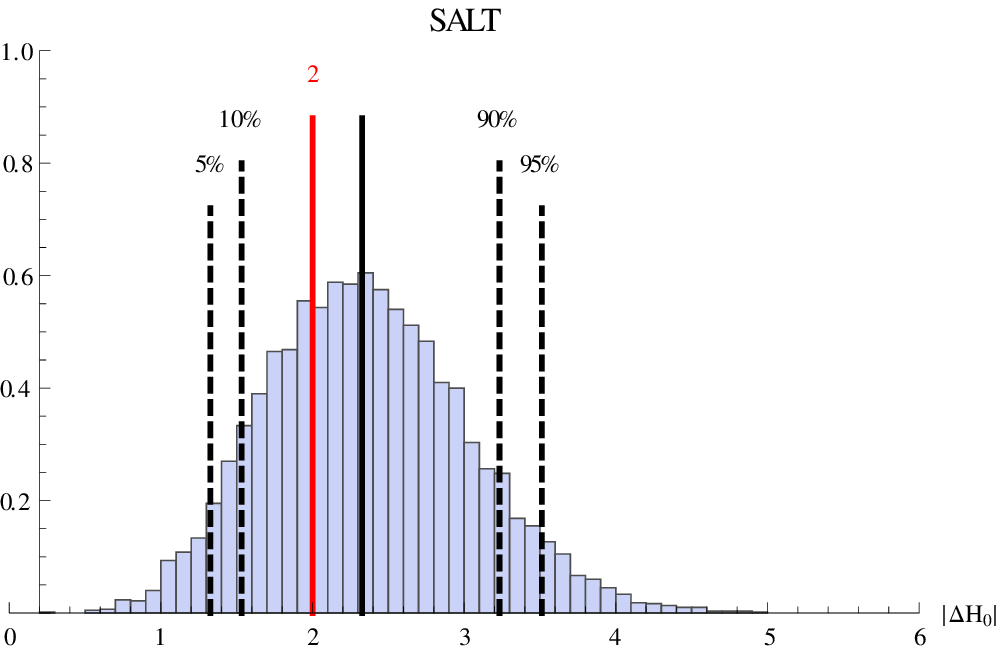}
	\hspace{0.06\linewidth}
	\includegraphics[width=0.47\linewidth]{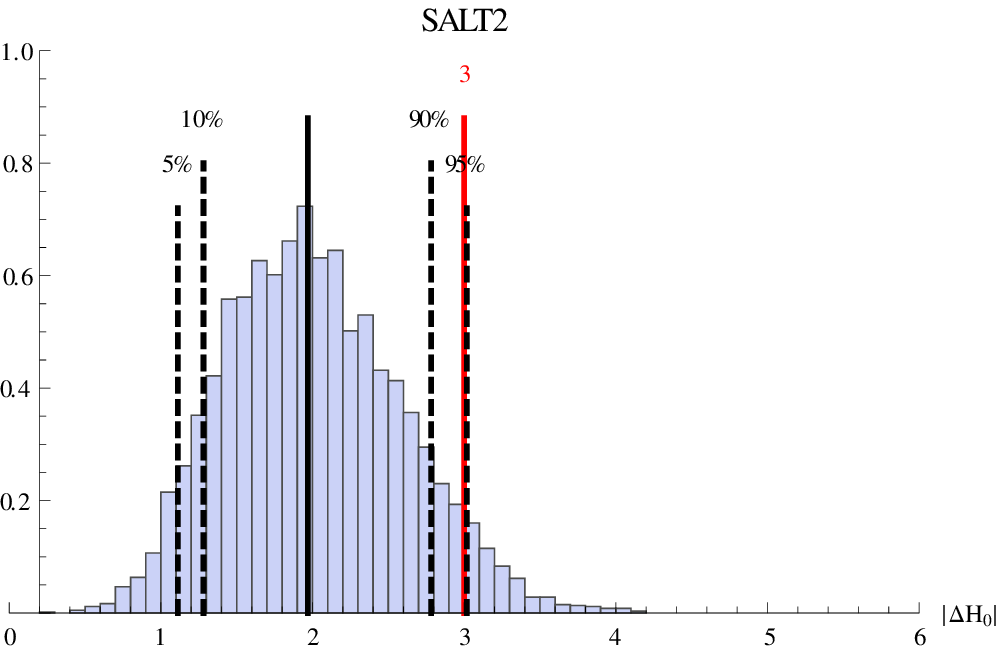}
	\caption{4 Histograms showing the maximal hemispherical asymmetry in the Hubble rate $H_0$ 
	  after scrambling 16000 times the positions of SNe in the MLCS2k2 (1.7) 
	  (top left), MLCS2k2 (3.1) (top right), SALT (bottom left) and SALT II (bottom right) data sets. Corresponding to Fig. \ref{fig:HN-HS_while_q0_fixed}, the 
	  deceleration parameter $q_0=-0.601$ is fixed. The red line denotes the value one gets by analysing the original data from \citet{Hicken}, the dashed lines 
	  denote the quantiles.}
	\label{fig:4HistoH0q0}
\end{figure*}

In our first approach we do not relate a SN from the Constitution set to its original position 
from the CBAT list, but relocate it at the position of another random SN, 
that is also part of the same sample of SNe. We repeat the procedures described in the previous section for 16000 realisations, which were generated by scrambling the 
positions. Thereafter, we again maximise the asymmetries in $H_N-H_S$, but as to save computing time, we do not calculate the asymmetry in every grid point but first 
evaluate the hemispherical Hubble asymmetry at 10 different random positions and then use the maximum of these as a start position $(\ell,b)_0$ for the
simulated annealing method described in \citet{Haggstrom}. In this method a Metropolis chain is simulated by determining the Hubble asymmetry on a randomly picked 
neighbouring
point $(\ell,b)_{n+1}=(\ell_{n}\pm 1^\circ,b_{n})$ or $(\ell,b)_{n+1}=(\ell_{n},b_{n}\pm 1^\circ)$. The new position $(\ell,b)_{n+1}$ is either accepted 
with a probability 
\begin{equation}
 p_{\rm acc}=\min\left\lbrace\frac{\exp\left(\frac{1}{T_{n+1}}\frac{H_N-H_S}{H_N+H_S}(\ell,b)_{n+1}\right)}
 {\exp\left(\frac{1}{T_{n+1}}\frac{H_N-H_S}{H_N+H_S}(\ell,b)_n\right)},1\right\rbrace
\end{equation}
 or rejected with $1-p_{\rm acc}$, i.e.
 $(\ell,b)_{n+1}=(\ell,b)_{n}$. The algorithm converges to the position $\lim_{n\rightarrow \infty}(\ell,b)_{n}
=(\ell,b)_{\rm max}$ at which the hemispherical Hubble asymmetry is maximised, as long as $T_n\propto\frac{1}{\ln(n)}$ \citep{Geman}. The algorithm is trained to find
the same Hubble asymmetry values as in Table~\ref{tab:Results_q0_fixed} with the same accuracy. 
Out of 10 runs, we identified the maximum number of steps needed to obtain 
these values and we chose a slightly higher number of steps, after which we truncate the Metropolis chain and save the highest expansion asymmetry appearing therein.

{
Assuming a perfectly isotropic Hubble expansion, any deviation from hemispherical symmetry would be due to selection effects. Thus scrambling the original data
yields an estimate of the distribution one would expect for an isotropic universe, which is measured 
with the true selection function. Moreover, no model has to be adopted
for this test.
}

The results are visualised in the histograms of Fig. \ref{fig:4HistoH0q0}. There, one can see again a 
distinction between the results for the SALT and all the other data 
samples: the maximum asymmetry in $H_0$ for the SALT data set is a fair realisation of this Monte Carlo 
procedure, whilst between 90 and 95 \% of the maxima of all Monte 
Carlo 
realisations are smaller than the asymmetry in the Hubble diagrams for the other light-curve fitters.

\subsection{Distance modulus Monte Carlo}

\begin{figure*}
	\includegraphics[width=0.47\linewidth]{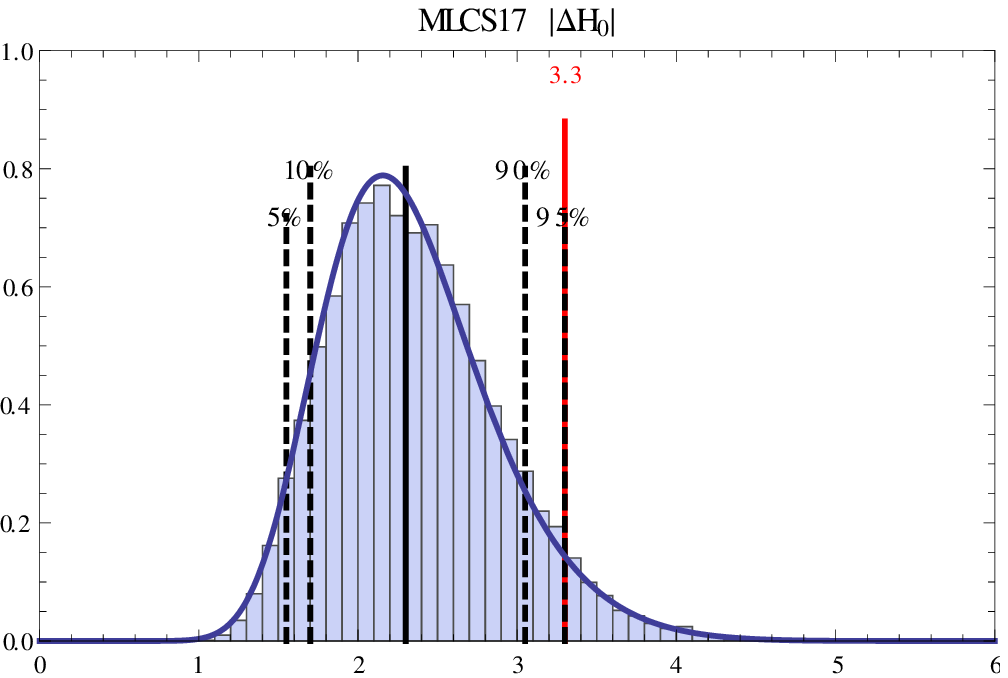}
	\hspace{0.06\linewidth}
	\includegraphics[width=0.47\linewidth]{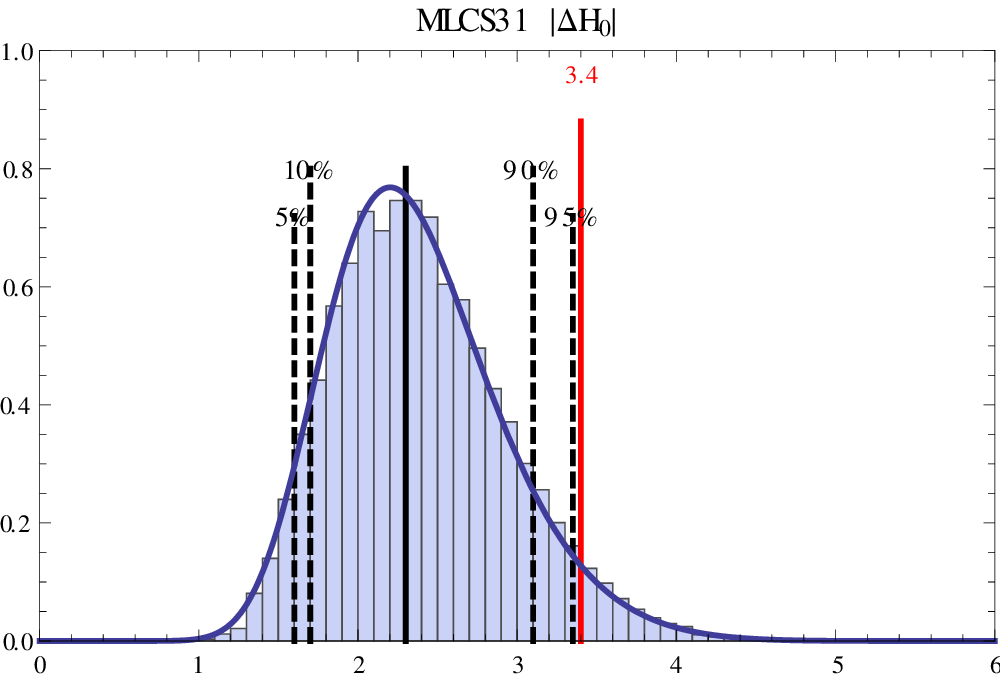}
	\includegraphics[width=0.47\linewidth]{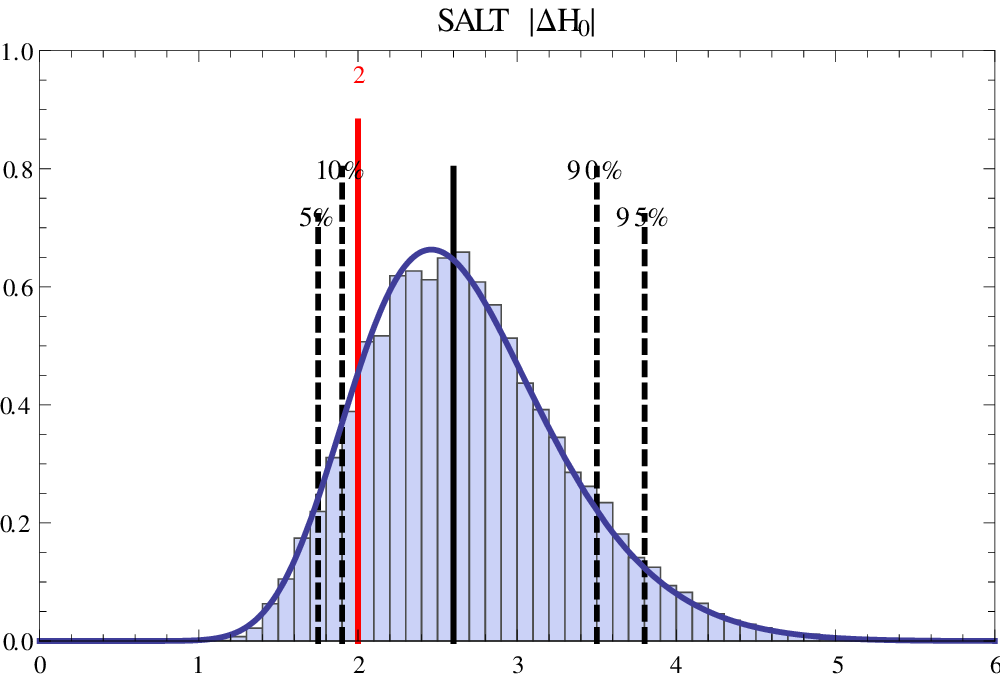}
	\hspace{0.06\linewidth}
	\includegraphics[width=0.47\linewidth]{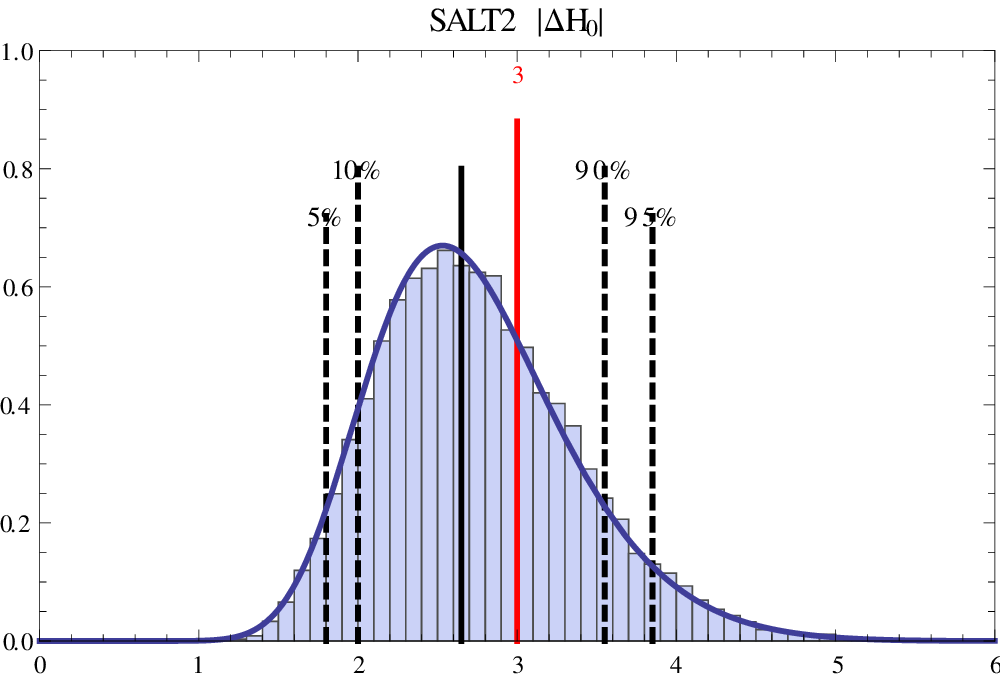}
	\caption{Histograms of the maximum asymmetry in $H_0$ found in 16000 different realisations of the distance modulus Monte Carlo, applied on the 
	  MLCS2k2 1.7, 
	  MLCS2k2 3.1, SALT and SALT II data sample, and the respective probability density functions of a generalised extreme value distribution. The distance moduli
	  in the simulated data are normally distributed around their theoretically calculated values with the original measurement errors $\sigma_{\mu_i}$ as standard 
	  deviation.}
	\label{fig:DistModH0Histo}
\end{figure*}

As we filled in the previous analysis observed positions with observed SNe, the simulated data can also be affected by an anisotropic distribution of observed SNe. 
Another way to check whether the asymmetries are significant is to repeat the following procedure:
\begin{enumerate}
 \item We generate a data set, in which we keep the positions and redshifts of each SN, but the distance modulus $\mu$ is calculated from its redshift $z$ 
       according to (\ref{eq:dist_mod}), where we assume the WMAP $\Lambda$CDM values $q_0=-0.601$ and $H_0=71.0\;\mathrm{km/s/Mpc}$ from \citet{Larson}.
 \item We simulate ``observations'' with uncertainties of measurement as a Gaussian distribution $\mathcal N (\mu, \sigma)$ , where the expectation value $\mu$ 
       equals the previously calculated distance modulus $\mu$ and the standard deviation $\sigma$ equals the measurement uncertainty $\sigma_{\mu_i}$ of the distance 
       modulus of the $i$th SN in the original data set.
 \item We maximise again the anisotropy $H_N-H_S$ in the simulated data set by dint of the simulated annealing method \citep{Haggstrom}.
\end{enumerate}

As the Hubble parameter is normally distributed, and thence the difference $H_N-H_S$ as well, we expect the maximum asymmetry in $H_0$ to obey a generalised 
extreme value distribution\footnote{Let $X_1,X_2,X_3,...$ be independent and identically distributed (iid) random variables. Let $F$ be the underlying cumulative 
distribution 
function. Then the probability, that the maximum of the iid random variables is smaller than $x$, is ${\bf P}\left(\max\left(X_1,...,X_n\right)\leq x\right)=F^n(x)$.
Suppose there exists a sequence of constants $a_n>0$, and $b_n\in\mathbb{R}$, such that $\frac{\max\left(X_1,X_2,...,X_n\right)-b_n}{a_n}$ has a non-degenerate limit 
distribution $\lim_{n\rightarrow\infty}F^n\left(a_n x+b_n\right)=G(x)$. Such a distribution $G$ is called an extreme value distribution (see e.g. \citet{deHaan}).}, due to the Theorem of 
\citet{Fisher} and \citet{Gnedenko}. The probability 
density function of which with location parameter $\mu$, scale 
parameter $\sigma$ and shape parameter $\xi\neq 0$ is given by 
\begin{equation}
 f(x; \mu, \sigma, \xi)=\frac{1}{\sigma}\left[1+\xi\left(\frac{x-\mu}{\sigma}\right)\right]^{-1/\xi-1}e^{-\left[1+\xi\left(\frac{x-\mu}{\sigma}\right)\right]^{1/\xi}}.
\end{equation}
The maximum likelihood parameter estimates can be found in Table~\ref{tab:GEVmaxlklhd}. For an infinite number of realisations, we would expect $\xi=0$, also
known as Gumbel distribution. As shown in Table~\ref{tab:GEVmaxlklhd}, $\left\vert\xi\right\vert\ll 1$.

\begin{table}
 \begin{tabular}{*{4}{l}r}
  \hline
  \noalign{\smallskip}
  fitter & $\mu$ & $\sigma$ & $\xi$ & 1-CDF$\left(\frac{H_N-H_S}{H_N+H_S}\right)_{max}$\\
  \noalign{\smallskip}
  \hline
  \noalign{\smallskip}
  MLCS2k2(1.7) & 2.11 & 0.47 & -0.10 & 5.1\%\\
  MLCS2k2(3.1) & 2.14 & 0.48 & -0.12 & 4.4\%\\
  SALT & 2.40 & 0.56 & -0.11 & 86.2\%\\
  SALT II & 2.47 & 0.55 & -0.11 & 30.3\%\\
  \noalign{\smallskip}
  \hline
 \end{tabular}
 \caption{Maximum likelihood parameter estimates $(\mu, \sigma, \xi)$ obtained from the distance modulus Monte-Carlo with different light-curve fitters (cf. fig 
  \ref{fig:DistModH0Histo}). The last column
  contains the probability of measuring higher Hubble asymmetries than the original ones according to the fitted distribution.}
 \label{tab:GEVmaxlklhd}
\end{table}

The resulting histograms (Fig. \ref{fig:DistModH0Histo}) show statistical significance for the anisotropies found in the MLCS2k2 data, which are higher than 95\% of the 
anisotropies in the simulated data, but they show no significance for the SALT and SALT II asymmetries. The distribution for MLCS2k2 is narrower than the others, because
the SALT sample contains less SNe and because SALT II yields larger errors for the distance moduli.

As to study how improving measurements of $\mu_i$ affects the usefulness of our hemispheric asymmetry test, we repeat 5000 distance modulus Monte Carlos, but we 
replace
 $\sigma_\mu$ by $\sigma_\mu/2$. {We can so show,} that if the measurement uncertainty of the
distance moduli of every single supernova can be halved, the Hubble asymmetries are expected to be much smaller. Fitting again a generalised extreme value 
distribution, {the maximum likelihood location parameters would amount 1.09 km/s/Mpc and 1.24 km/s/Mpc for MLCS2k2 (3.1) and 
SALT II respectively.} The observed values would lie at {the} $1-10^{-6}$ and $1-4\times 10^{-5}$ 
quantiles for {the MLCS2k2 (3.1) and SALT II light-curve fitters.} We hope, this will encourage observers to further reduce measurement errors, as it also 
improves this test considerably.

\subsection{Comparison with cosmic variance}

{
Finally, we want to determine the fluctuations of the hemispherical asymmetry
around its average, for a random orientation of the hemispheres. To calculate
these fluctuations, we pick 10,000 random positions ($\ell$,$b$) in the sky.
Each position so determines the north pole of a random sphere. We split this
sphere into its northern and southern hemisphere and calculate from the Constitution set the hemispherical
Hubble asymmetry $\frac{H_N-H_S}{H_N+H_S}$ of this randomly oriented sphere. The
variance of the Hubble asymmetry for the 10,000 spheres is then the desired
fluctuation.}

{To sample the random north pole positions ($\ell$,$b$), we employ a restriction:
If we chose them from the whole sky [$\ell$ uniformly distributed, ${\rm
pdf}(b)=\frac{1}{2}\cos(b)$], for each position where we take ($\ell$,$b$) to be
the north pole of our random sphere, we would also have a random sphere in the
sample where ($\ell$,$b$) is the south pole, because we chose the position
($\pi+\ell$,$-b$) as a north pole. So in order for the average to be the
typical asymmetry for random orientation, we have to restrict the choice of
random north pole positions ($\ell$,$b$) to one hemisphere of the sky.}

{In order not to miss a direction of high anisotropy, we do three runs for
orthogonal directions in the sky that we take as the poles of the hemispheres on
which we sample our random positions ($\ell$,$b$). In the first run, we generate
only positions ($\ell$,$b$) on the northern galactic hemisphere, in the second
run positions are restricted to the western hemisphere and in the third
repetition the generated ($\ell$,$b$) are located on the hemisphere towards the
galactic centre.}

{Table~\ref{tab:meandeltaH0} compares the empirical root mean square fluctuations
$\sigma_{emp}$ of the hemispherical Hubble asymmetry determined by this
prescription with the fluctuations in the Hubble asymmetry due to cosmic
variance $\sigma_{A}$. Under the assumption that each configuration of SN in the
sphere obtained by the random choice of its north pole position represents a
realisation of the stochastic process that generates the SN distribution,
$\sigma_{emp}$ should give an estimate of $\sigma_{A}$.
Table~\ref{tab:meandeltaH0} shows, that their values indeed are quite
comparable. Only the empirical values obtained with MLCS2k2 (1.7) and SALT II
are slightly higher than expected from this reasoning.
}

\begin{table}
	\begin{tabular}{p{0.225\linewidth}*{3}{p{0.2\linewidth}}}
		\hline
		\noalign{\smallskip}
		fitter & hemisphere & $\sigma_{emp}\left(\frac{H_N-H_S}{H_N+H_S}\right)$ & $\sigma_{A}$\\
		\noalign{\smallskip}
		\hline
		\noalign{\smallskip}
		MLCS2k2 (1.7) & GC & 0.8\% & {0.7\%} \\
			      & NH & 1.0\% &  {0.7\%} \\
			      & WH & 1.0\% &  {0.7\%} \\
		MLCS2k2 (3.1) & GC & 0.7\% &  {0.7\%} \\
			      & NH & 0.8\% &  {0.7\%} \\
			      & WH & 0.8\% &  {0.7\%} \\
		SALT & GC & 0.6\% &  {0.7\%} \\
		     & NH & 0.5\% &  {0.7\%} \\
		     & WH & 0.6\% &  {0.7\%} \\
		SALT II & GC & 0.8\% &  {0.6\%} \\
			& NH & 0.9\% &  {0.6\%} \\
			& WH & 0.7\% &  {0.6\%} \\
		\noalign{\smallskip}
		\hline
	\end{tabular}	
	\caption{Empirical standard deviation $\sigma_{emp}$ of the distribution of $\frac{H_N-H_S}{H_N+H_S}$ evaluated on random positions on the northern (NH) and 
	  western (WH) hemisphere
	  as well as on the hemisphere towards the galactic centre (GC) compared with expected fluctuations $\sigma_{A}$ 
	  due to cosmic variance.}
	\label{tab:meandeltaH0}
\end{table}

\section{Systematics}
\label{sec:systematics}
\subsection{Equatorial hemispheres}
\begin{figure}
	\centering
	\includegraphics[width=\linewidth]{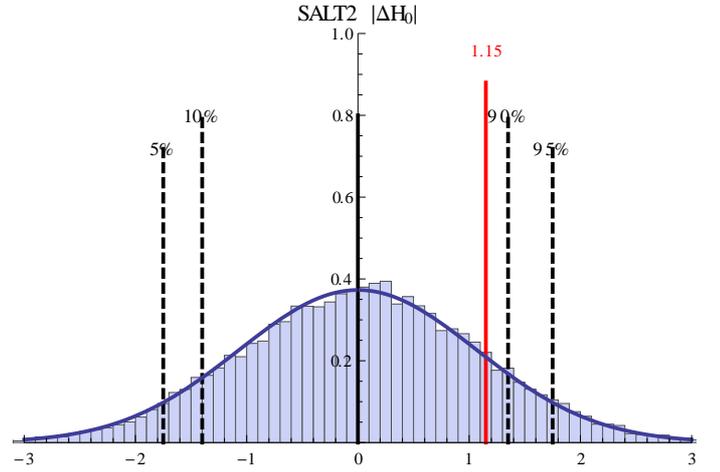}
	\caption{Histogram with the resulting differences between the Hubble rate on the northern equatorial hemisphere $H_N$ and the southern equatorial hemisphere 
	$H_S$ obtained from 16000 runs of the distance modulus Monte Carlo. The red line indicates the measured value $H_N-H_S=1.15\mathrm{\frac{km}{s\cdot Mpc}}$ obtained 
	from the SALT II data set. We also plotted a normal distribution $\mathcal{N}(\mu, \sigma)$ with $\mu=0.00$ and 
	$\sigma=1.07$.}
	\label{fig:Equatorial}
\end{figure}

In this section we are concerned with the question how unlikely is the asymmetry observed towards a fixed direction on the sky. Therefore, we carry out again the 
distance modulus Monte Carlo, but now we do not maximise the anisotropy, but we calculate the differences in the Hubble parameters of the equatorial hemispheres. 
The results all look like those obtained from SALT II (Fig. \ref{fig:Equatorial}), where the measured value $H_N-H_S=1.15\;\mathrm{\frac{km}{s\cdot Mpc}}\;\hat{=}\;
\frac{H_N-H_S}{H_N+H_S}=0.88\%$ lies within the 90\% confidence interval of the resulting distribution. The corresponding values for MLCS2k2 1.7 and 3.1, as well as for 
SALT are 
$1.00\;\mathrm{\frac{km}{s\cdot Mpc}}$, $1.25\;\mathrm{\frac{km}{s\cdot Mpc}}$ and $-0.45\;\mathrm{\frac{km}{s\cdot Mpc}}$ or accordingly 0.76\%, 0.96\% and -0.34\%. 
These values are all smaller than the cosmic variance. Thence and because the values lie within their corresponding 90\% confidence intervals, this test does not support 
that the 
observation strongly depend on Earth related effects. 

\subsection{Number asymmetry}

\begin{figure*}
	\includegraphics[width=0.5\linewidth]{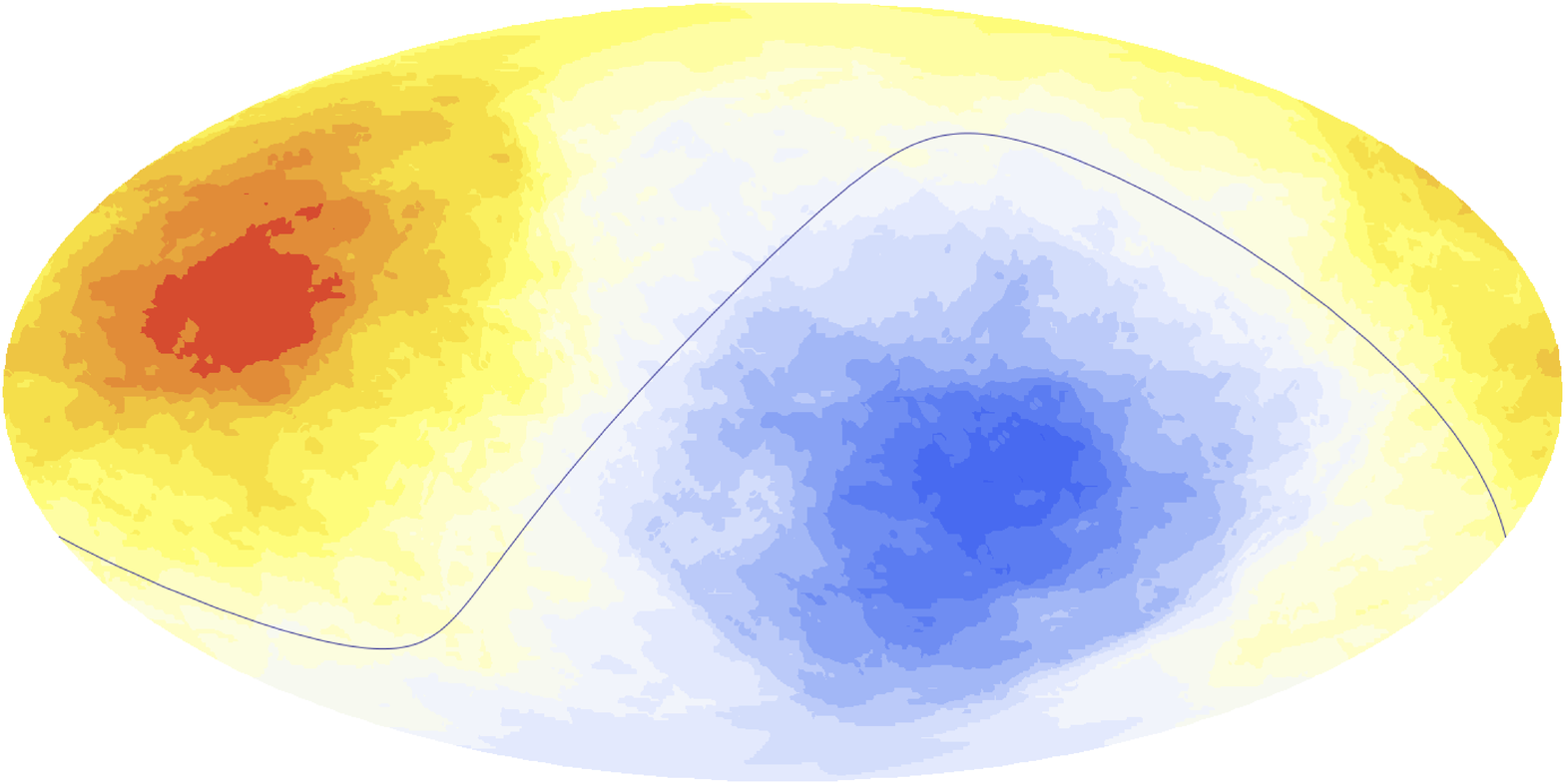}
	\includegraphics[width=0.5\linewidth]{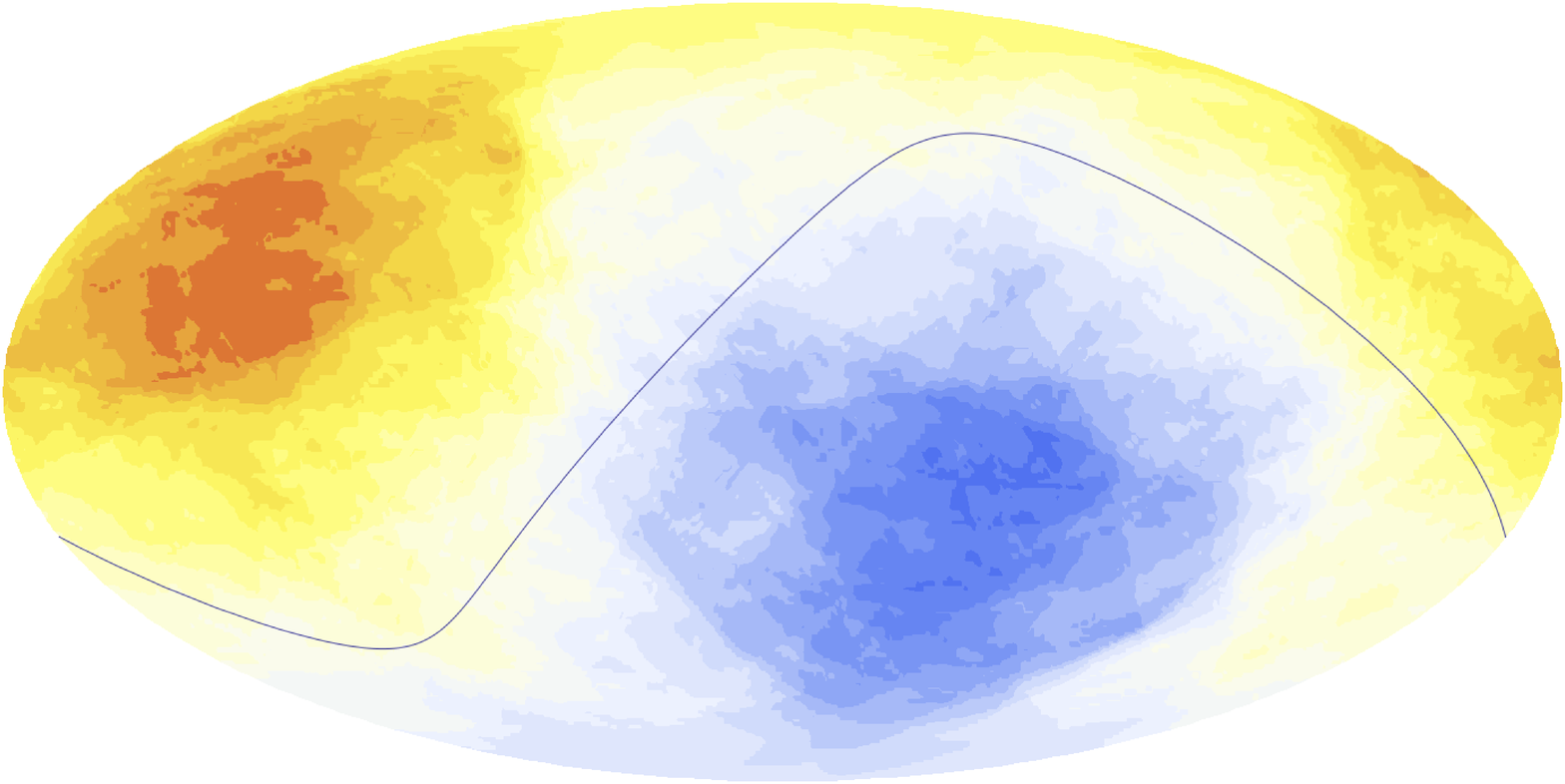}
	\includegraphics[width=0.5\linewidth]{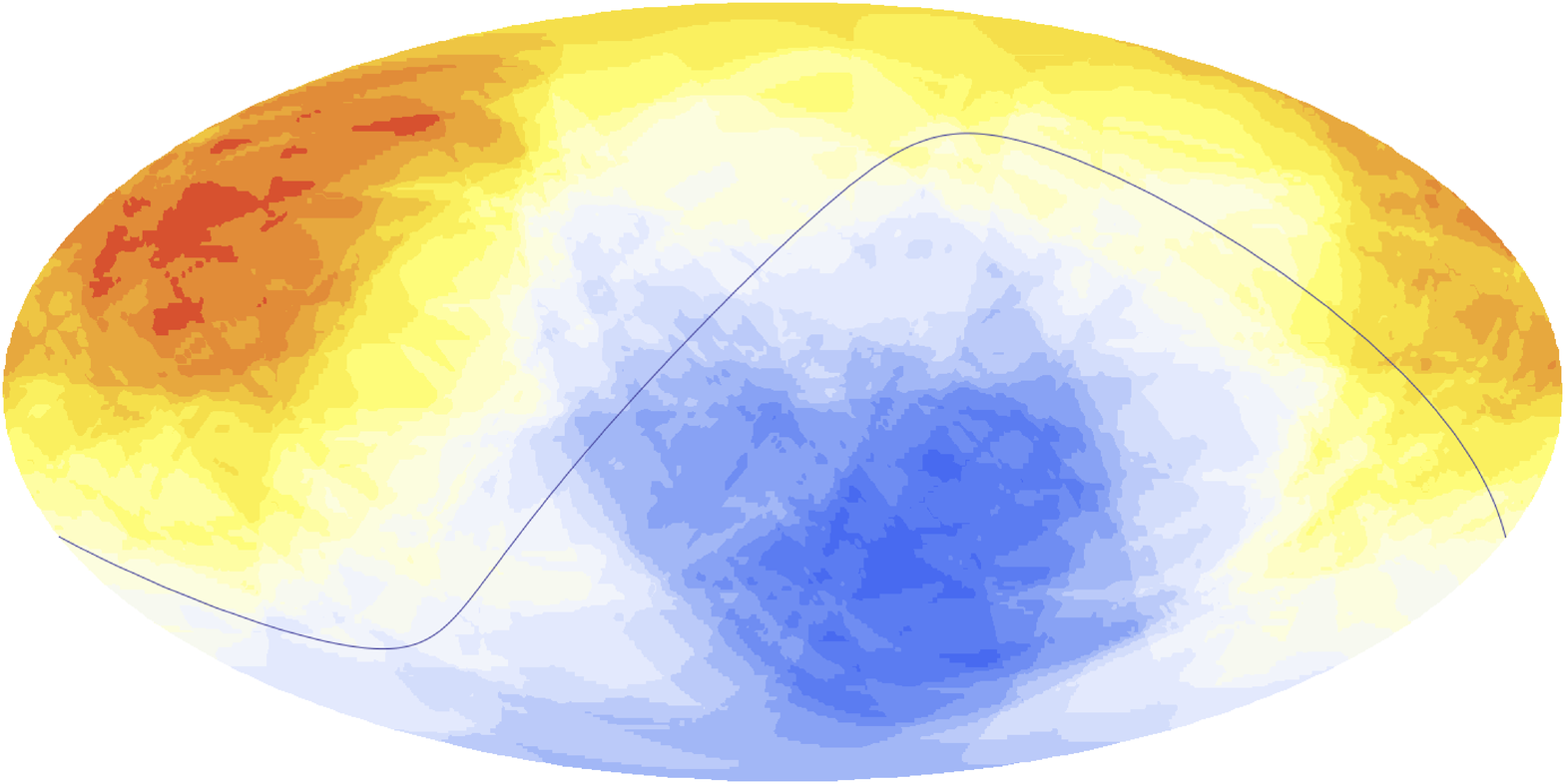}
	\includegraphics[width=0.5\linewidth]{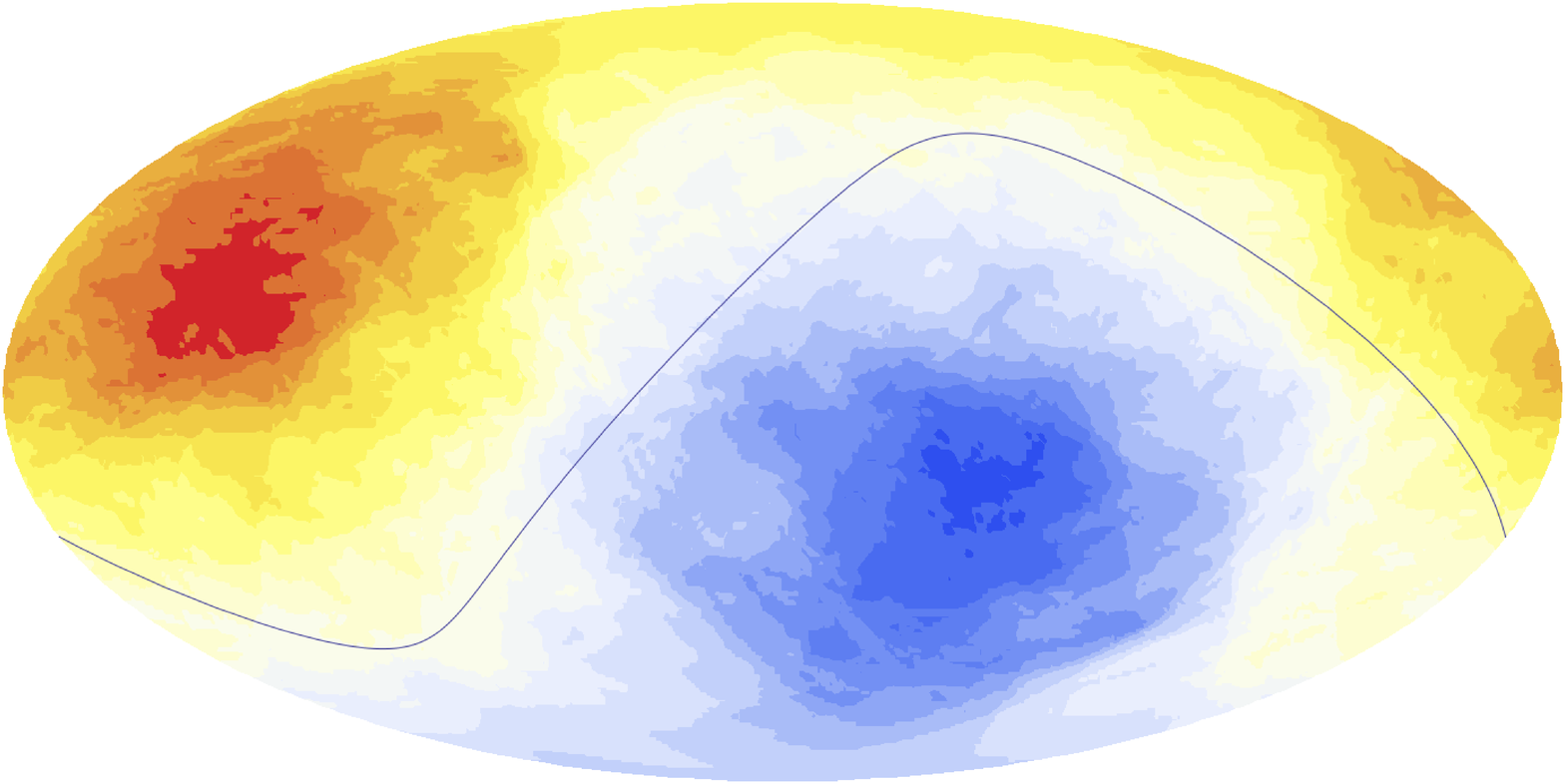}
	\caption{Hemispherical asymmetry $\frac{N_N-N_S}{N_N+N_S}$ in the number $N_i$ of SNe on the Hemisphere labelled with $i$. The plots are adjusted in the 
	  same way as Fig. \ref{fig:HN-HS_while_q0_fixed}. The maximum value, which defines the scale, is 0.52.}
	\label{fig:Number}
\end{figure*}

As already mentioned, the asymmetries in $H_0$ can be due to incomplete SNe Ia observations in some directions. Therefore, we plotted the difference between the number 
of SNe on both hemispheres similarly to Fig. \ref{fig:HN-HS_while_q0_fixed}. These plots (Fig. \ref{fig:Number}) show anisotropies in 
$\frac{N_N-N_S}{N_N+N_S}$, but they do not agree with any asymmetry map in Fig. \ref{fig:HN-HS_while_q0_fixed}. The asymmetry in the SN distribution 
is minimal near the Earth's equator and maximal approximately at its poles. Also the white region, in which the number asymmetry is zero, follows the equatorial 
equator. This is not the case in Fig.~\ref{fig:HN-HS_while_q0_fixed}.

Although the maximal asymmetries in the number of SNe is around 50\% in all data sets, the number asymmetries for the hemispheres that give the maximal expansion
anisotropies (cf. Table \ref{tab:Results_q0_fixed}) are much smaller for MLCS2k2 and SALT II (MLCS2k2 1.7: -4.8\%, MLCS2k2 3.1: -20.2\%, SALT II: -10.6\%). Only SALT
manifests a comparably high number asymmetry between the hemispheres of maximum expansion asymmetry, that is to say 46.9\%, even though it is the most isotropic subsample
in terms of Hubble asymmetry.

For that reason we cannot think of the incompleteness of the SN data being the main explanation of asymmetries found in the Hubble expansion.

\subsection{Quality of fit}

Another possible systematic effect could be, that the quality of our fits is different in different directions. We have verified, that the asymmetries in 
$\frac{\chi^2}{\mathrm{dof}}$ are
also not aligned with the asymmetries in $H_0$.

\subsection{Dust asymmetry}

\begin{figure}
	\centering
	\includegraphics[width=\linewidth]{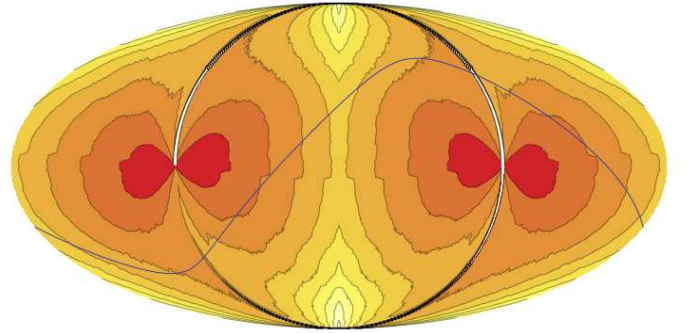}
	\caption{Hemispherical asymmetry in the antenna temperature of the foreground dust extracted from \citet{Schlegel}. The maximum asymmetry is 0.14 mK.}
	\label{fig:Schlegel}
\end{figure}

We assume, that the Universe is isotropic on large scales. The foreground, like dust within our galaxy, is not isotropic and therefore a possible explanation of 
apparent anisotropic expansion.\\
For the comparison with galactic dust, we brought the Schlegel dust map \citep{Schlegel} into the same form as our other maps. The result can be seen in 
Fig. \ref{fig:Schlegel}. This map has no similarities with Fig. \ref{fig:HN-HS_while_q0_fixed}, thus, we can also exclude dust as a reason for apparent anisotropic 
cosmic expansion.

\subsection{Shape asymmetry}

\begin{figure*}
	\includegraphics[width=0.5\linewidth]{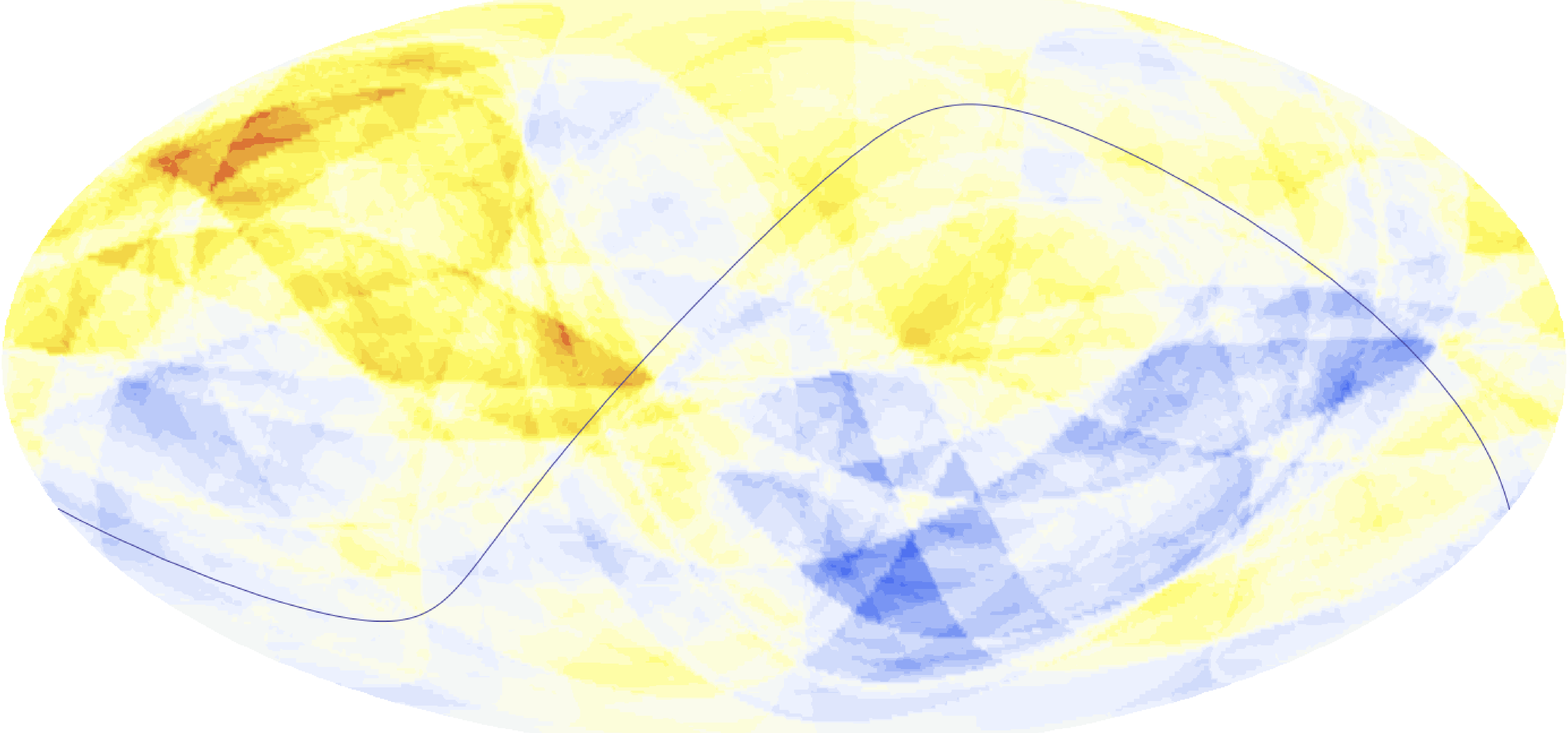}
	\includegraphics[width=0.5\linewidth]{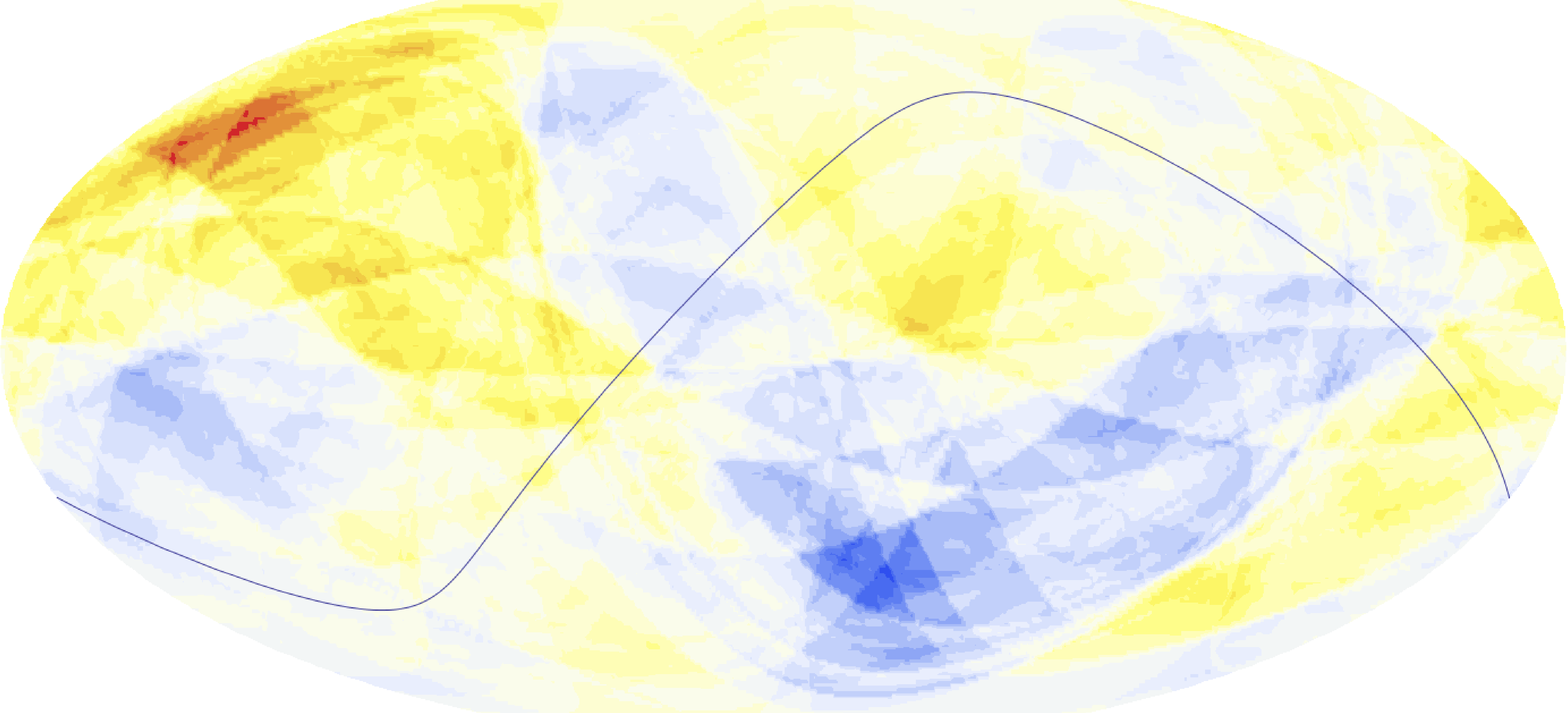}
	\caption{Asymmetry $\frac{\Delta_N-\Delta_S}{\Delta_N+\Delta_S}$ of the shape parameter $\Delta$. The scaling is defined by the maximum value of 0.57.
                 The plot on the left is for MLCS2k2(1.7), the other one for MLCS2k2(3.7).}
	\label{fig:Shape}
\end{figure*}

The MLCS2k2 light-curve fitter uses the light-curve shape to remove the influence of dust from the distance measurement \citep{Riess}. As \citet{Hicken} provides 
the shape parameter of each SN, we also plot the weighted difference $\frac{\Delta_N-\Delta_S}{\Delta_N+\Delta_S}$ between the average shape parameters $\Delta$ 
of each hemisphere (see Fig.~\ref{fig:Shape}). These plots also do not reproduce the shapes of the asymmetry map of $H_0$ (Fig.~\ref{fig:HN-HS_while_q0_fixed}).

\subsection{Colour asymmetry}

\begin{figure*}
	\includegraphics[width=0.5\linewidth]{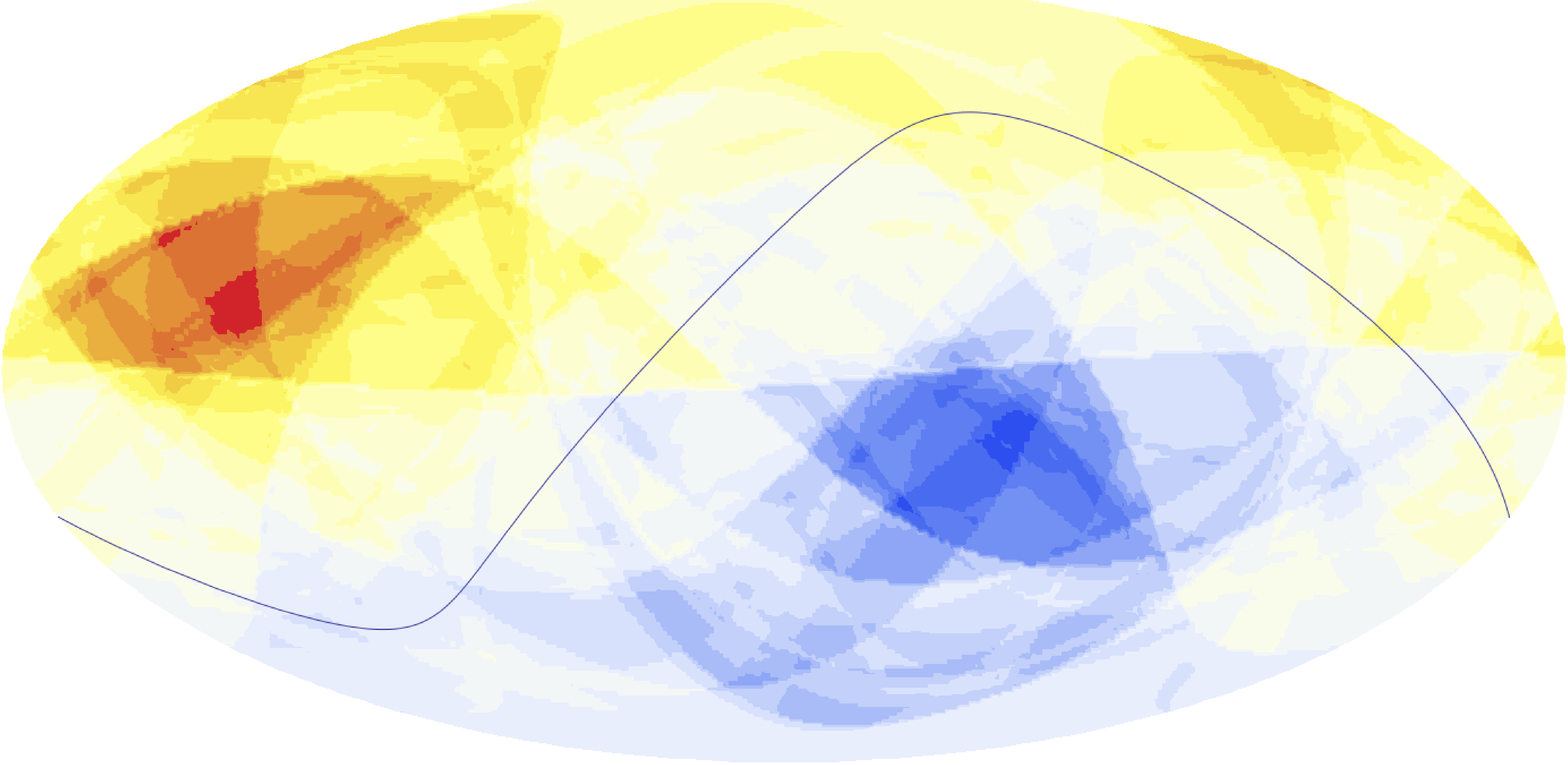}
	\includegraphics[width=0.5\linewidth]{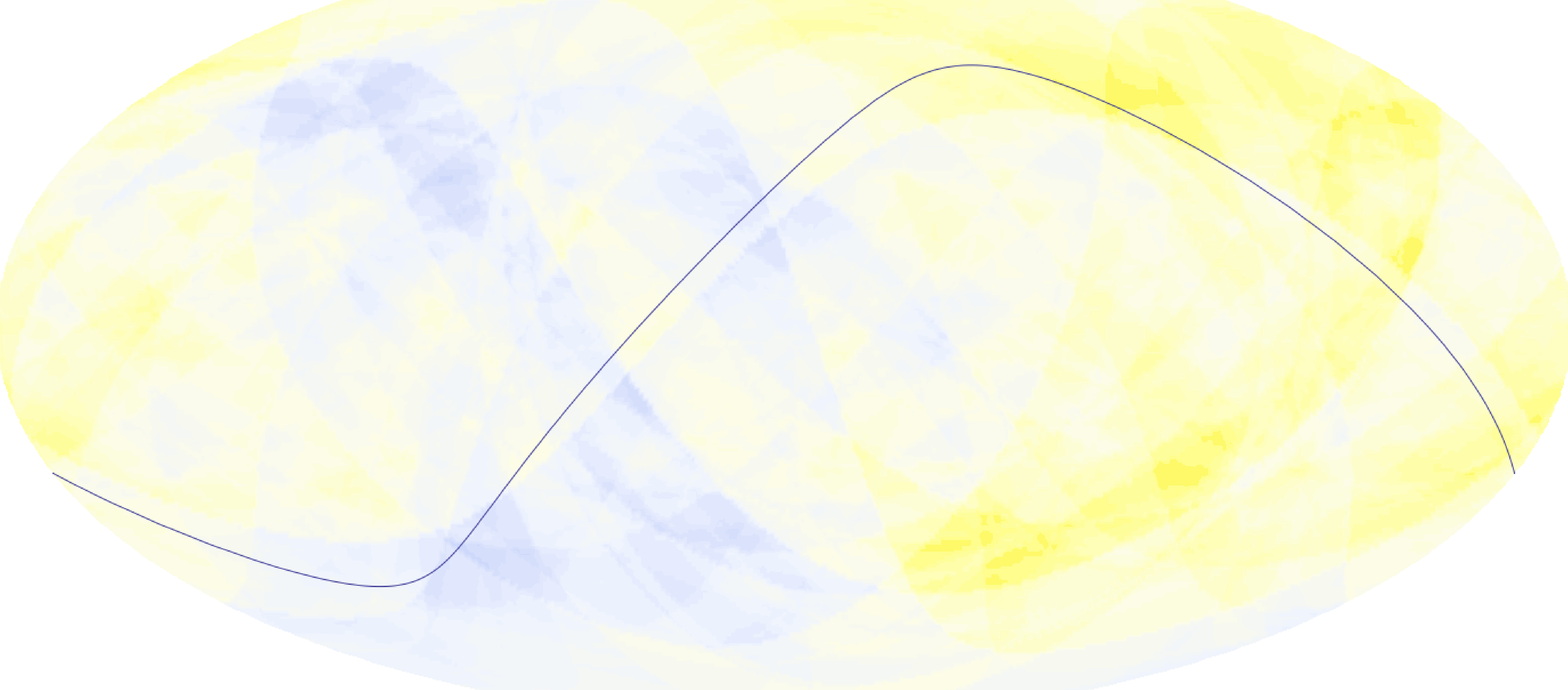}
	\caption{Asymmetry $\frac{c_N-c_S}{c_N+c_S}$ of the colour parameter. The maximum value amounts to 0.64. The plot on the left hand side shows the colour 
                 asymmetry in the SALT data, the right one in SALT II.}
	\label{fig:Colour}
\end{figure*}

SALT and SALT II use a colour parameter to get rid of dust \citep{Guy05, Guy07}. We tested the asymmetries in the colour parameter in the same way as those in 
the shape parameter. The resulting maps are shown in Fig. \ref{fig:Colour}. The SALT data set shows directions in which the colour parameter is obviously asymmetric, as 
against irregular asymmetries of the colour parameter in the SALT II data.

\section{Limits on expansion asymmetry}
\label{sec:results}
As to find limits on the Hubble expansion asymmetry of the local Universe, {we draw
realisations from the SN magnitudes and redshifts given the reported measurements with errors.} 
We then compute the asymmetry of the original direction of maximum asymmetry given in Table~\ref{tab:Results_q0_fixed}. After 5400 realisations, we 
enumerate the
95\% quantile $H_{95}$, which can be regarded as an upper limit on the expansion asymmetry, as well as the 5\% 
quantile $H_5$. In an isotropic universe, all measurements
of the Hubble asymmetry should agree with $\Delta H/H=0$ at every position. Hence, a positive $H_5$ cannot be explained by statistical fluctuations alone. 

$H_5$ and $H_{95}$ can be found for every light-curve fitter in Table~\ref{tab:limits}. SALT is 
consistent with an isotropic local Universe, whilst the other light-curve
fitters provide evidence for real anisotropies. MLCS2k2 (3.1) features an $H_5$, which is even 
comparably high as our expectations for typical fluctuations due to large scale 
structure.

\begin{table}
 \begin{tabular}{lll}
  \hline
    \noalign{\smallskip}
    fitter & 5\% quantile & 95\% quantile\\
    \noalign{\smallskip}
    \hline
    \noalign{\smallskip}
    MLCS2k2 (1.7) & 0.8\% & 3.4\%\\
    MLCS2k2 (3.1) & 1.1\% & 3.8\%\\
    SALT & 0.0\% & 3.0\%\\
    SALT II & 0.7\% & 3.7\%\\
  \hline
 \end{tabular}
 \caption{Lower and upper limits on the hemispherical expansion asymmetry}
 \label{tab:limits}
\end{table}

In Fig. \ref{fig:Conclusion}, we plot the contours of the 5\% quantiles of the MLCS2k2(3.1) and SALT II as error contours of the direction of maximum Hubble asymmetry. Both
data sets agree with the expansion being fastest into the direction of the WMAP cold spot I reported by \citet{Bennett}.

The $H_{95}$ values provide upper limits on the anisotropy of the Hubble expansion. The least stringent upper limit is found from MLCS2k2 (3.1), namely
$\frac{H_N-H_S}{H_N+H_S}<0.038$ at 95\% C.L.

\section{Conclusion}
\label{sec:concl}

\begin{figure*}
	\includegraphics[width=\linewidth]{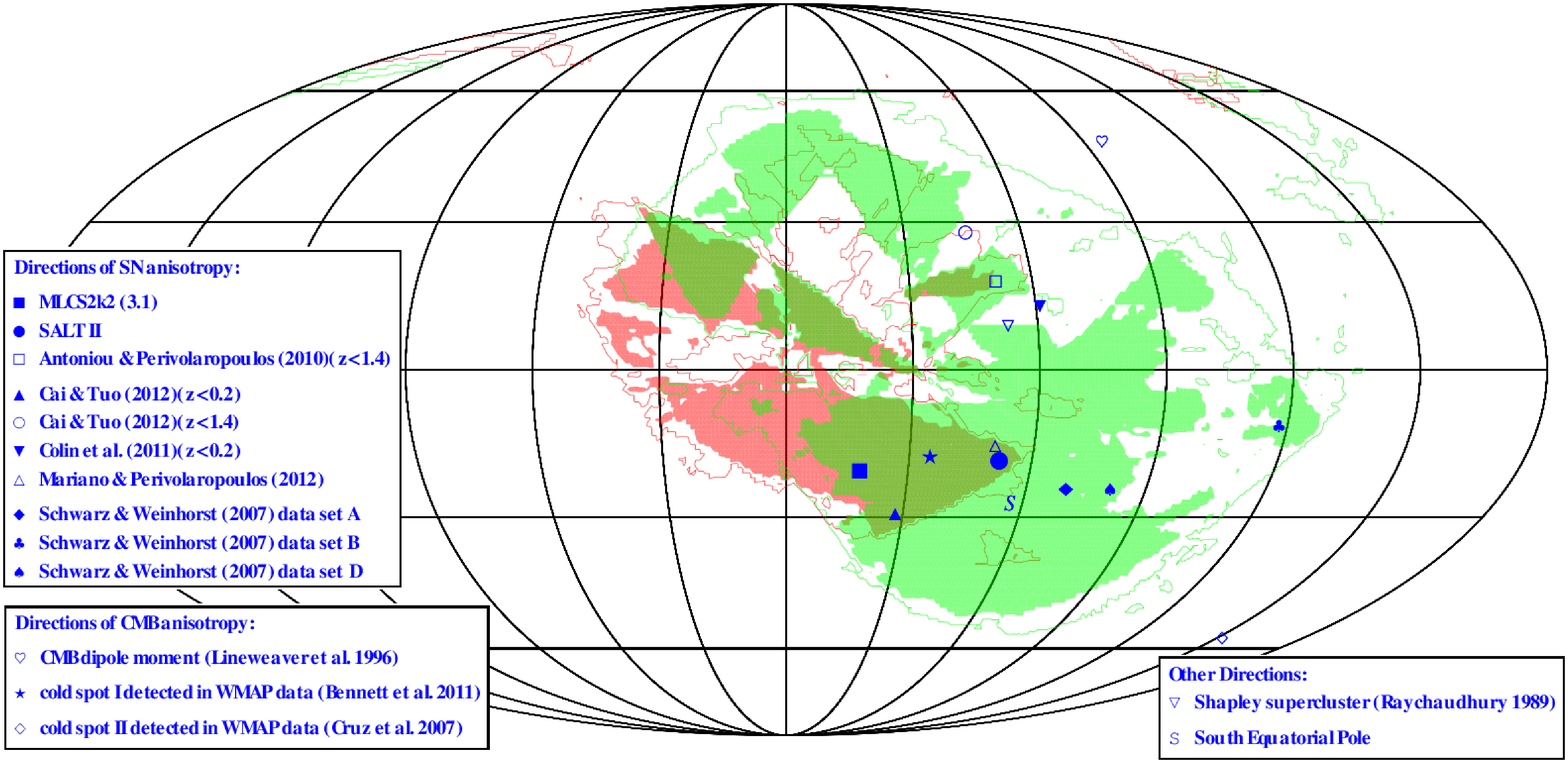}
	\caption{95 \% confidence level contours of the SALT II (green lines) and MLCS2k2(3.1) (red lines) maximum asymmetry directions (MLCS2k2(3.1) 
		$\blacksquare$, SALT II {\large$\bullet$}
		) as well as 90\% confidence level 
		contours (filled green and filled red 
		respectively, filled dark green where they coincide) compared with directions of 
		anisotropies in SN data from \citet{Antoniou} ($z<1.4$) $\square$, 
		\citet{Cai} ($z<0.2$ $\blacktriangle$, $z<1.4$ $\circ$), \citet{Colin} ($z<0.2$ $\blacktriangledown$), 
		\citet{Mariano} $\vartriangle$ and \citet{Schwarz} (data set A $\blacklozenge$, B $\clubsuit$, D $\spadesuit$), and with 
		the direction of the CMB dipole moment $\heartsuit$ \citep{Lineweaver}, the Shapley supercluster $\triangledown$ \citep{Raychaudhury}, 
		the cold spots I 
		$\bigstar$ \citep{Bennett} and II detected in WMAP data $\Diamond$ \citep{Cruz}, and the South Equatorial Pole $S$.}
	\label{fig:Conclusion}
\end{figure*}

We analyse hemispherical asymmetries $\Delta H/H=(H_N-H_S)/(H_N+H_S)$ in the Hubble expansion rate at redshifts $z<0.2$ in SN data of the Constitution set published by 
\citet{Hicken}. The observed 
hemispherical asymmetry
of the Hubble expansion is in agreement with the order of magnitude that can be expected by cosmic variance in a $\Lambda$CDM universe, but is inconsistent
with the assumption of perfect isotropy and homogeneity.

However, the Hubble asymmetry in the direction of
maximal asymmetry as identified by the MLCS2k2 light curve fitter, is at 95\% C.L. larger than in a
random realisation of distance modulus observations.
All light-curve fitters, bar SALT, establish their respective
asymmetric directions at 90\% C.L. in terms of data scrambling. Fig. \ref{fig:Conclusion} shows their agreement with directions reported by \citet{Cai, Mariano} and 
\citet{Schwarz}, which are in the vicinity to the WMAP cold spot I \citep{Bennett}. \citet{Inoue} explained observed large-angle CMB anomalies with a pair of local 
dust-filled voids.
These would cause both a cold spot in CMB data and fluctuations in the locally measured Hubble constant as large as 2-4\%. Our measured Hubble anisotropy of 
$\Delta H/H \sim 0.026$ agrees with this finding. Despite the fact that it is significant, we so conclude that the anisotropy does not contradict global isotropy, 
because 
firstly it can be explained by local structure, secondly we only analysed local SNe and thirdly it matches our predictions of typical fluctuations in a $\Lambda$CDM 
background. 

Admittedly, our test is not sensitive enough to see if the 1\% asymmetry detected for random hemispheres is due to
$\Lambda$CDM-fluctuations of the local structure. So it should be repeated, when a full sky
survey comprising approximately 2000 SNe becomes available, or when the errors of individual SNe could be reduced by about a factor of 2. But nevertheless 
we can constrain the expansion 
asymmetry of the local Universe to be less than 3.8\% at 95\% C.L. using most conservatively the results obtained with the MLCS2k2 (3.1) light-curve fitter.

\begin{acknowledgements}
We thank Marek Kowalski for valuable discussions.
We acknowledge the use of the List of Supernovae provided by the IAU Central Bureau for Astronomical Telegrams 
({\tt http://www.cbat.eps.harvard.edu/lists/Supernovae.html}).
We acknowledge financial support from Deutsche Forschungsgemeinschaft under grants IRTG 881 and RTG 1620.
MS is supported by the South African Square Kilometre Array Project.
\end{acknowledgements}

\bibliographystyle{aa} 
\bibliography{120806} 

\end{document}